\documentclass{aa}  

\usepackage{graphicx}
\usepackage{txfonts}
\usepackage{hyperref}

\hypersetup{
    colorlinks=true,
    linkcolor=blue,
    urlcolor=blue,
    citecolor = blue,
}

\begin{document}

   \title{The AMBRE Project: Spectrum normalisation influence on \\ Mg abundances in the metal-rich Galactic disc}

   \author{P. Santos-Peral
          \inst{1}
          \and
          A. Recio-Blanco\inst{1}
          \and
          P. de Laverny\inst{1}
          \and
          E. Fernández-Alvar\inst{1}
          \and
          C. Ordenovic\inst{1}
          }

   \institute{\inst{1} Laboratoire Lagrange (UMR7293), Université de Nice Sophia Antipolis, CNRS, Observatoire de la Côte d’Azur, BP 4229, F-
    06304 Nice Cedex 04, France\\
              \email{psantos@oca.eu}
         %\and
             }

   \date{Received January 17 2020 / Accepted June 2 2020}

% \abstract{}{}{}{}{} 
% 5 {} token are mandatory
 
  \abstract
   {The abundance of $\alpha$-elements provides an important fossil signature in Galactic archaeology to trace the chemical evolution of the different disc populations. High-precision chemical abundances are crucial to improving our understanding of the chemodynamical properties present in the Galaxy. However, deriving precise abundance estimations in the metal-rich disc ([M/H] > 0 dex) is still challenging. }
   {The aim of this paper is to analyse different error sources affecting magnesium abundance estimations from optical spectra of metal-rich stars.}
   {We derived Mg abundances for 87522 high-resolution spectra of 2210 solar neighbourhood stars from the AMBRE Project, and selected the 1172 best parametrised stars with more than four repeated spectra. For this purpose, the GAUGUIN automated abundance estimation procedure was employed.}
   {The normalisation procedure has a strong impact on the derived abundances, with a clear dependence on the stellar type and the line intensity. For non-saturated lines, the optimal wavelength domain for the local continuum placement should be evaluated using a goodness-of-fit criterion, allowing mask-size dependence with the spectral type. Moreover, for strong saturated lines, applying a narrow normalisation window reduces the parameter-dependent biases of the abundance estimate, increasing the line-to-line abundance precision. In addition, working at large spectral resolutions always leads to better results than at lower ones. The resulting improvement in the abundance precision makes it possible to observe both a clear thin-thick disc chemical distinction and a decreasing trend in the magnesium abundance even at supersolar metallicities.}
   {In the era of precise kinematical and dynamical data, optimising the normalisation procedures implemented for large spectroscopic stellar surveys would provide a significant improvement to our understanding of the chemodynamical patterns of Galactic populations.}

   \keywords{Stars:abundances --
                The Galaxy: disc --
                Methods: data analysis
               }

    \maketitle
%
%________________________________________________________________

\section{Introduction}
Disentangling the chemodynamical signatures present in the disc's stellar populations is essential to unveiling the formation and evolution of the Milky Way. Since the first thin-thick disc identification \citep{yoshii1982, gilmore1983}, chemical signatures (preserved in FGK-type stars' atmospheres) have been suggested as the best criteria to differenciate between Galactic stellar populations. Numerous studies \citep[e.g.][]{vardan2012,alejandra2014, bensby2014, wojno2016, ivanyuk2017, buder2019, haywood2019, hayden2020} have characterised these two Galactic components in the solar neighbourhood. \par

In particular, the $\alpha$-elements abundance (e.g. O, Mg, Si, S, Ca, Ti) has been widely analysed to chemically disentangle the Galactic thin-thick disc populations \citep{fuhrmann2011, vardan2012, alejandra2014,hayden2017}. The observed thick disc has been reported to be [$\alpha$/Fe]-enhanced relative to the thin disc for most metallicities, unveiling distinct chemical evolution histories in both disc components. The abundance of [$\alpha$/Fe] is used as a good chronological proxy. This is due to the timescale delay between core-collapse supernovae (Type II SNe) of the most massive stars (\textit{M $\gtrsim$ 8M$_\odot$}), which enrich the ISM with $\alpha$-elements predominantly, and Type Ia SNe, which release mainly iron-peak elements \citep{matteucci1986}.  However, both high- and low-$\alpha$ sequences seem to overlap at supersolar metallicites, showing a flat trend for most $\alpha$-process elements (not expected by chemical evolution models) being impossible to chemically identify to which stellar population they belong. In addition, different features of the disc [$\alpha$/Fe] abundances as the intermediate $\alpha$ populations at high metallicities \citep{vardan2012, sarunas2017} and the gap between these stars and the high-$\alpha$ metal-poor population \citep{vardan2012, gazzano2013}, are still matter of debate. \par

In the solar neighbourhood and beyond, several spectroscopic stellar surveys have provided valuable chemical information and constraints, such as SEGUE \citep[][R$\sim$1800]{segue}, LAMOST \citep[][R$\sim$1800]{lamost}, GES \citep{ges}, RAVE \citep[][R$\sim$7500]{rave}, APOGEE \citep[][R$\sim$22500]{apogee}, and GALAH \citep[][R$\sim$28000]{galah}. Nevertheless, the flattening of the [$\alpha$/Fe] abundances at
supersolar metallicities is a common feature of many different studies. For instance, \citet{anders2014} and \citet{hayden2015} showed the [$\alpha$/Fe] vs [Fe/H] plane across the Milky Way for a large sample
of giant stars from APOGEE. They find that both low- and high-[$\alpha$/Fe] sequences decrease with [Fe/H], merging and showing a flattened trend at supersolar metallicities. Recently, \citet{buder2019} found similar signatures over the sample of dwarfs and turn-off stars from GALAH DR2, showing a remarkable agreement with the results of \citet{vardan2012}. Similarly, \citet{sarunas2017} defined both Galactic discs chemically finding the flattening trend in the [Mg/Fe] ratio for stars with metallicity [Fe/H] > -0.2 dex in a dwarf star sample from AMBRE data.  \par

Theoretically, two-infall chemical evolution models \citep{chiappini1997, romano2010} predict a steeper slope in the metal-rich regime ([Fe/H] > 0.0 dex), underestimating the measurement of Mg abundances with respect to the observed flat trend. More recent models \citep[e.g.][]{kubryk2015} seem to reproduce this flattening including stellar radial migration, introduced by \citet{radialmix}, through churning and blurring.  Observational evidence of radial migration has also been noticed in several studies, using data from different surveys \citep{georges2015, hayden2017, feltzing2020}. Updated chemical evolution models developed by \citet{grisoni2017, grisoni2018} conclude also that other mechanisms are needed, in addition to the inside-out formation scenario, to reproduce the flattened trend in the Galactic disc at supersolar metallicites. \par
   
On the basis of these apparent discrepancies, the study of chemical signatures requires the best possible precision and accuracy in the abundance measurement. Precise abundances are mandatory to detect stellar populations that differ in their elemental abundances from each other \citep{lindegren2013}. Magnesium is probably the most representative and commonly used $\alpha$-element \citep[c.f.][]{carrera2019}. It is known to have a high number of measurable spectral lines in optical spectra. In addition, the ratio of the Mg abundance with respect to iron, [Mg/Fe], shows a large absolute separation of the Galactic thick-thin disc populations, along with a smaller scatter and a shallower trend with temperature and metallicity \citep{brewer2016, bergemann2017, ivanyuk2017, buder2019}, making this element possibly the best tracer. \par
   
A detailed exploration of possible error sources is crucial to interpreting the reality of the observed chemical signatures in the Galactic stellar populations and the resulting implications on chemodynamical relations (such as the contribution of radial migration in the solar neighbourhood or the use of [Mg/Fe] as a good age proxy), which are mainly constrained by the abundance precision. The main issues concerning the determination of high-precision abundances are characterised by the need for both high signal-to-noise ratio (S/N) and spectral resolution, and predominantly by the definition of continuum to normalise the observed spectral data. The latter issue can be responsible for the largest fraction of the uncertainty in the abundance estimations, which is still complex for cool metal-rich stars due to the high presence of blended and molecular lines \citep[as reviewed by][]{nissen2018,biblia2019}. In particular, the continuum normalisation is not fully optimised for different stellar types in large spectroscopic stellar surveys. \par

In this paper, we present a detailed spectroscopic analysis of the Mg abundance estimation for a sample of 2210 FGK-type stars in the solar neighbourhood observed and parametrised at high spectral resolution within the context of the AMBRE Project \citep{patrick2013}. We point out that we refer to the observed high- and low-[Mg/Fe] sequences as thick and thin disc populations hereafter, although we have not applied any kinematic selection to the sample. The paper is organised as follows. In Sect. \ref{data}, we introduce the observational data sample used in this work. The automatic abundance estimation method is described in Sect. \ref{method}. We present the sources of error caused by the normalisation procedure in Sect. \ref{normalization}. In Sect \ref{results}, we show the final derived [Mg/Fe] abundances of the sample. We conclude with a summary in Sect \ref{conclusions}.

%__________________________________________________________________

\section{The AMBRE:HARPS observational data sample} \label{data}
The AMBRE Project \citep{patrick2013} is a collaboration between the Observatoire de la Côte d’Azur (OCA) and the European Southern Observatory (ESO), of which the main goal is to determine the stellar atmospheric parameters (T$_{eff}$, log(g), [M/H], [$\alpha$/Fe]) of archived stellar spectra of the FEROS, HARPS, and UVES ESO spectrographs. The stellar parameters were derived by the multi-linear regression algorithm MATISSE \citep[MATrix Inversion for Spectrum SynthEsis,][]{alejandra2006}, developed at OCA and used in the Gaia RVS analysis pipeline \citep[\textit{Radial Velocity Spectrometer}, see][]{alejandra2016}, with the AMBRE grid of synthetic spectra \citep{patrick2012}. \par  

\begin{figure}[h]
\centering
\includegraphics[height=70mm, width=0.5\textwidth]{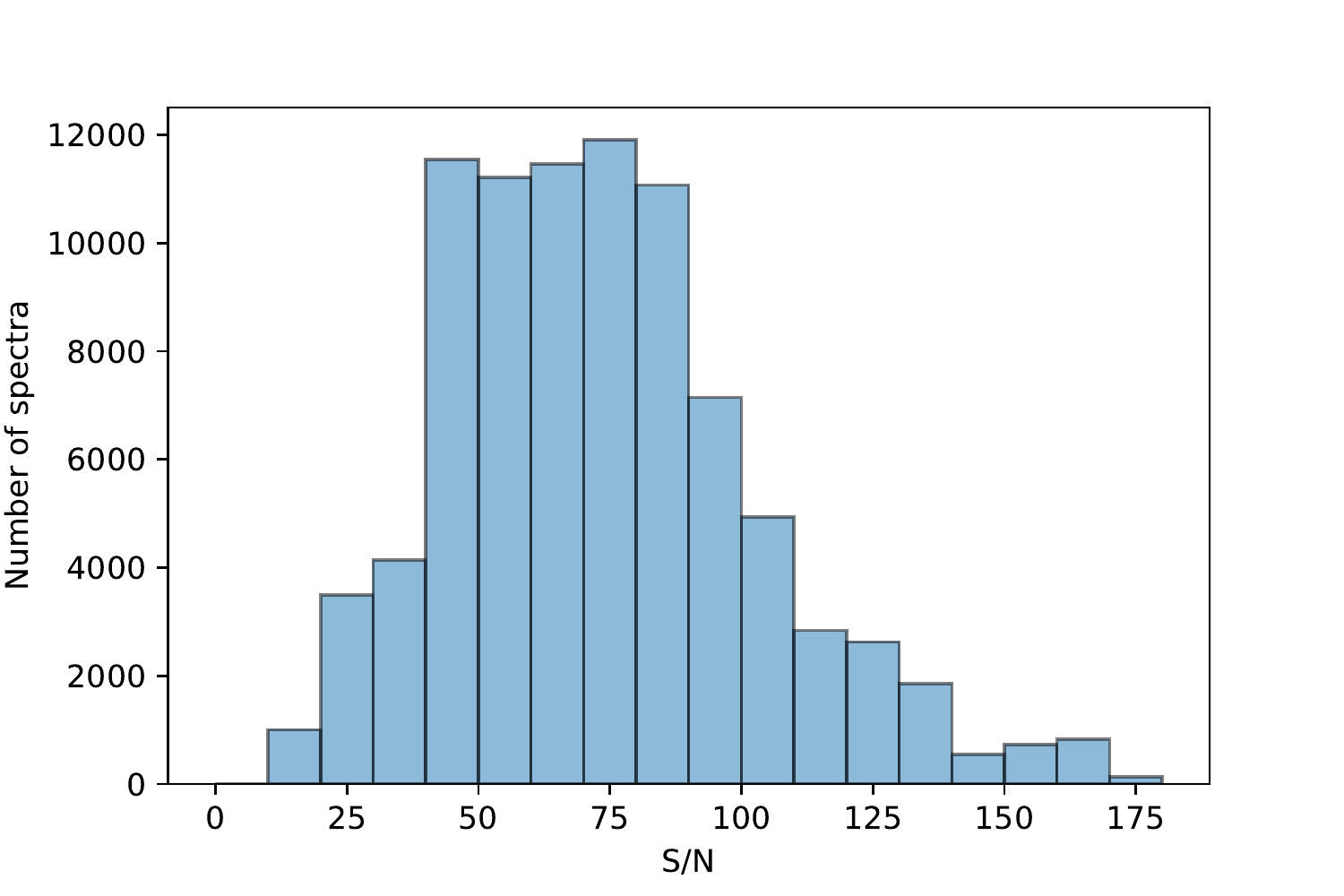}
\caption{Signal-to-noise ratio (S/N) for the spectra in our AMBRE:HARPS sample.}
\label{Fig:param_SNR}
\end{figure}

For the present paper, we derived [Mg/Fe] abundances over a sample of 87522 HARPS spectra\footnote{The AMBRE analysis of the HARPS spectra comprises the observations collected from October 2003 to October 2010 with the HARPS spectrograph at the 3.6m telescope at the La Silla Paranal Observatory, ESO (Chile).}, corresponding to 2210 stars. These spectra sample was selected according to the goodness of fit between the synthetic and the observed spectrum, keeping those that present a quality label of 0 or 1 \citep[see Table 3 of][]{DePascale2014}. The signal-to-noise ratio (S/N) distribution of the sample is shown in Fig. \ref{Fig:param_SNR}. We only considered the HARPS spectra sample due to the high spectral resolution \citep[R$\sim$115000,][]{mayor2003} and to avoid biases among the different spectrographs included in the AMBRE Project. 
\par

\vspace{-0.4cm}

\begin{figure*}[h]
\centering
\includegraphics[height=60mm, width=\textwidth]{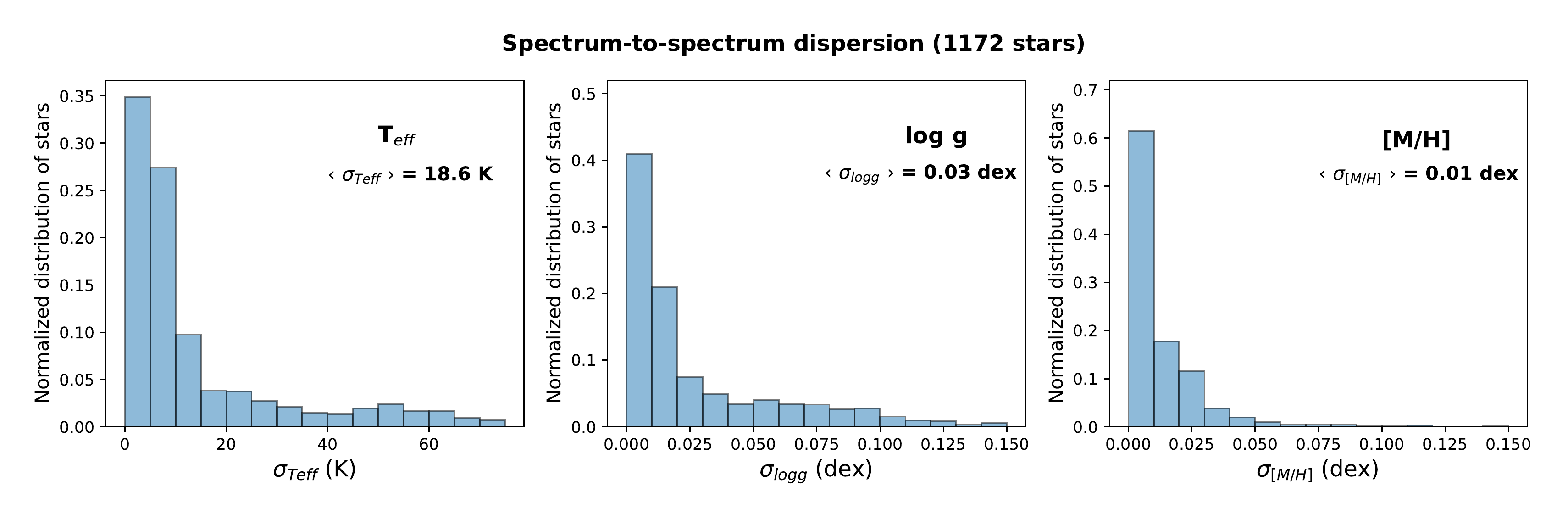}
\caption{Estimated dispersion of the selected stars ($\geq$ 4 repeats) on the effective temperature (left), the surface gravity (middle), and the global metallicity (right).}
\label{Fig:param_dispersion}
\end{figure*}

The HARPS sample contains a large number of repeated observations for some stars. A cross-match with the \textit{Gaia} DR2 catalogue \citep{gaia2018} allowed us to assign a \textit{Gaia} ID to each spectrum, identifying the different spectra of the same star. In order to be statistically significant and avoid spurious effects in single spectra, we analysed the stars with more than four observed spectra ($\geq$ 4 repeats), only selecting stars with T$_{eff}$ > 4700 K, as cooler stars could have larger errors in the parameters \citep[c.f. Fig. 12 in][where the cool main sequence flattens for metal-rich targets]{DePascale2014}. In addition, for each star we excluded spectra whose atmospheric parameters differ by more than two sigma from the mean value of said star. This allows us to discard possible mismatches and avoid the propagation of uncertainties on the atmospheric parameters to the stellar abundances. These different quality selections lead to a total of 76502 spectra, corresponding to 1172 stars. The estimated dispersion on the stellar atmospheric parameters from the different spectra of the same star are shown in Fig. \ref{Fig:param_dispersion}. The average dispersion on T$_{eff}$, log(g) and [M/H] are 18.6 K, 0.03 dex and 0.01 dex, respectively. \par

\begin{figure}
\centering
\includegraphics[height=75mm, width=0.5\textwidth]{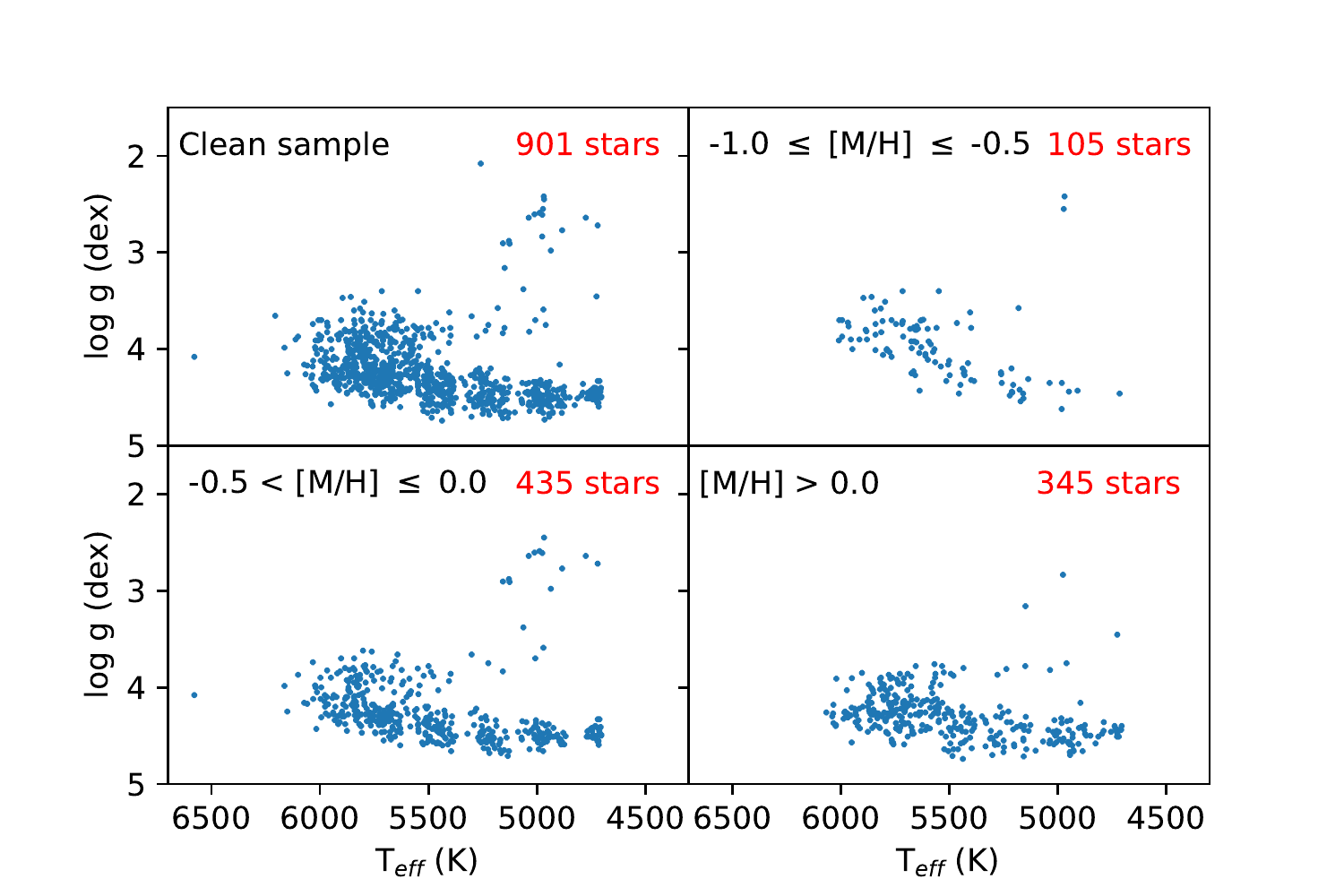}
\caption{HR diagram (in metallicity bins) of the selected AMBRE:HARPS stellar sample with FWHM$_{CCF}$ $\leq$ 8 km s$^{-1}$ and more than four observed spectra ($\geq$ 4 repeats).}
\label{Fig:HR_clean}
\end{figure}

To avoid any possible source of uncertainties from line-broadening, we only kept spectra with FWHM$_{CCF}$\footnote{Cross-correlation function between the observed spectra and the corresponding templates used for the radial velocity estimation.} $\leq$ 8 km s$^{-1}$. As a consequence, the selected AMBRE:HARPS sample is restricted to the stellar atmospheric parameters shown in Fig. \ref{Fig:HR_clean}, mostly dwarf stars cooler than 6200K.

%__________________________________________________________________

\section{Method} \label{method}
From the high-resolution observational spectra sample described in the previous section, we derived and analysed the [Mg/Fe] abundances using 9 Mg I spectral lines in the optical range automatically, via the optimisation method GAUGUIN \citep{Bijaoui2012, guiglion2016, alejandra2016}, and using a reference synthetic spectra grid produced in the framework of the Gaia-ESO Survey (GES) project \citep{ges}. GAUGUIN was part of the analysis pipeline of GES for the GIRAFFE spectra \citep{alejandra2014}, and is also used for Gaia RVS spectra \citep{alejandra2016}.  \par

The atmospheric parameters (T$_{eff}$, log(g), [M/H], [$\alpha$/Fe]) were used as an input (independently determined by the AMBRE Project, as described above). A first global normalisation procedure was iteratively attached to the parameter estimation. This iteration is described in  \citet{worley2012} and was performed by \citet{DePascale2014}. 
For the present abundance analysis, an initial global normalisation was applied, considering a large wavelength domain of 70\AA. In addition, a local normalisation around the considered spectral line was performed to optimise the continuum placement. Different widths of the local normalisation window were explored (c.f. Section \ref{normalization}). The local normalisation is not iterative. The details regarding the normalisation algorithm are described hereafter. 

Moreover, the radial velocity correction was performed using the accurate estimated provided by the ESO:HARPS reduction pipeline, except for a small proportion of the spectra with no HARPS radial velocity available, for which it was estimated by the AMBRE analysis procedure with similar precision \citep[see][]{worley2012, DePascale2014}. \par

\subsection{The normalisation procedure} \label{Normalisation}

\begin{figure}[h]
\centering
\includegraphics[height=70mm, width=0.52\textwidth]{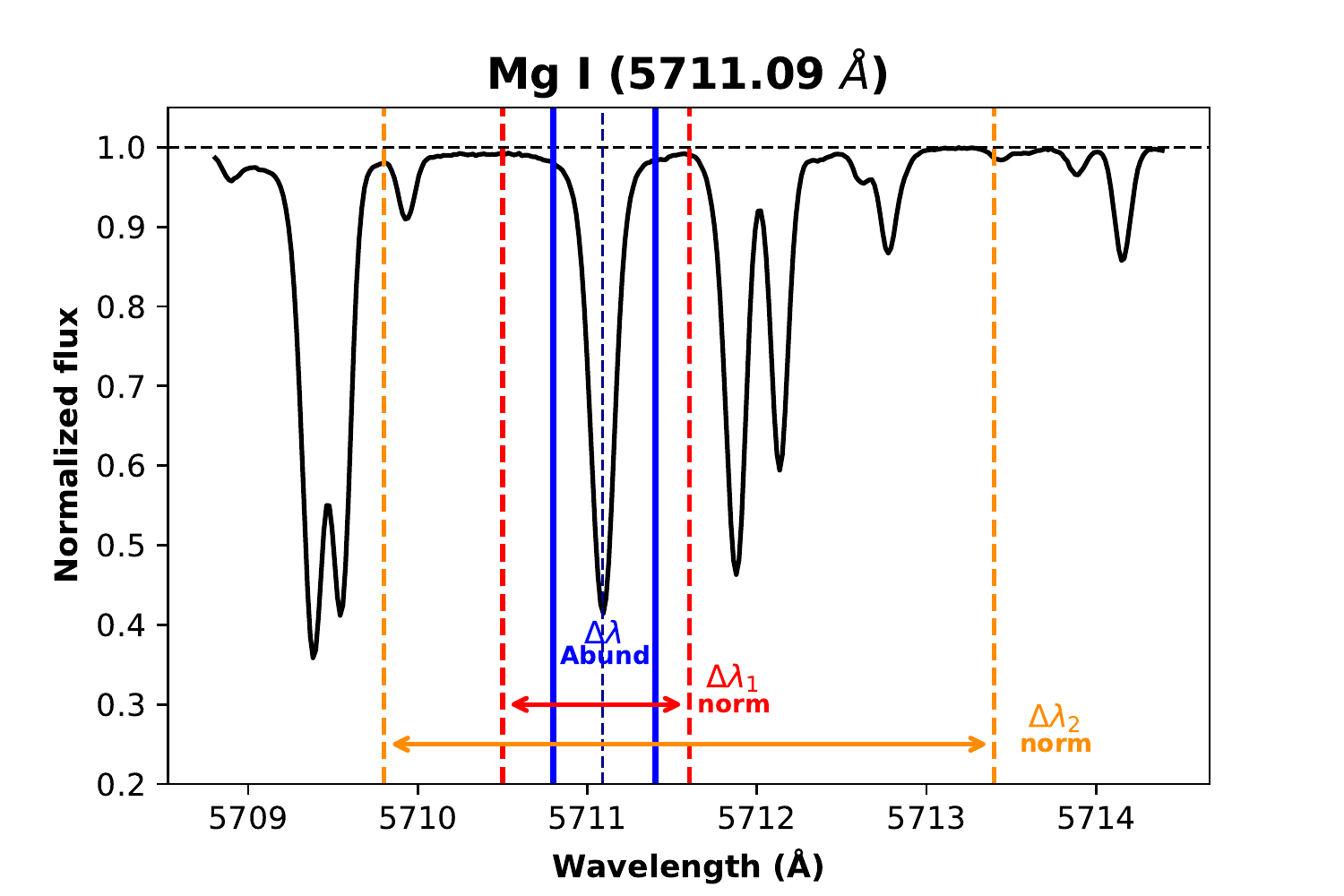}
\caption{Observed solar spectrum from HARPS (R=115000, asteroid reflection) around the line 5711.09 $\AA$. The adopted wavelength domain where the abundance is measured is delimited by blue vertical lines ($\Delta$$\lambda$$_{Abund}$ $\sim$ 0.5 $\AA$). Two different local normalisation intervals of 1$\AA$ ($\Delta\lambda_{1}$) and 4$\AA$ ($\Delta\lambda_{2}$) are shown with red and orange dashed vertical lines, respectively.}
\label{Fig:intervals_GAUGUIN}
\end{figure}

The observed spectrum flux was normalised over a given wavelength interval centred on the analysed line. For this purpose, the observed spectrum (O) was compared to an interpolated synthetic one (S) with the same atmospheric parameters. First, the most appropriate pixels of the residual (R = S/O) were selected using an iterative procedure implementing a linear fit to R followed by a $\sigma$-clipping. The clipping values vary from the first to the final iteration, starting with $\sigma_{-0.5}^{+5}$ and ending with $\sigma_{-1}^{+2}$. Then, a final residual was calculated (R$_{final}$ = S*$_{norm}$/O*$_{norm}$), where S*$_{norm}$ and O*$_{norm}$ are the synthetic and observed flux values in the previously selected pixels, applying an additional 0.2$\sigma$-clipping. Finally, the normalised spectrum was obtained after dividing the observed spectrum by a linear function resulting by the fit of R$_{final}$. No convolution was carried out during the normalisation procedure, so the original spectral resolution is conserved. As an example, Fig.~\ref{Fig:intervals_GAUGUIN} shows the normalised observed solar spectrum around the Mg line 5711.09 $\AA$.

\subsection{The GAUGUIN automated abundance estimation algorithm } \label{GAUGUIN}

The GAUGUIN code is a classical optimisation method based on a local linearisation around a given set of parameters from the reference synthetic spectrum, via linear interpolation of the derivatives. The abundance estimate is performed considering the spectral flux in a predefined wavelength window, which is always inside the defined local normalisation one. The abundance window ($\Delta$$\lambda$$_{Abund}$, c.f. Fig.~\ref{Fig:intervals_GAUGUIN}) was set around 0.5$\AA$ for non-saturated lines, and 2.5$\AA$ for strong saturated ones (see classification in Sect \ref{lines} below, $\sim$ 1.5-2 times the FWHM of each line in a solar-type star). We tested different local continuum intervals, always larger than these abundance estimation windows, to study the normalisation influence on the derived abundances for each type of line. It is worth noting that the local normalisation interval is not always perfectly symmetric around the analysed line. For some cases, due to contiguous strong absorption lines present on a particular side of the line, an asymmetric window is chosen in order to maximise the number of pixels close to the continuum level (see Appendix \ref{INTERVALS} for further details). For those particular configurations (only a few among the total analysed cases), the abundance window although included in the normalisation interval would be off centre. The observed abundance trends for these cases are consistent with the results obtained from symmetrically selected windows.
\par   
\par 

\begin{figure}[h]
\centering
\includegraphics[height=75mm, width=0.42\textwidth]{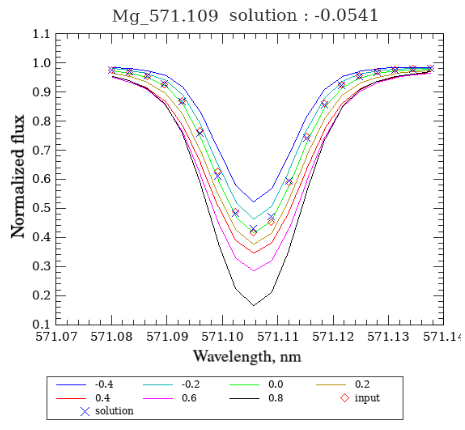}
\caption{Example of the fit carried out by the optimisation code GAUGUIN for the line 5711.09 $\AA$ of the observed solar spectrum. The normalised observed spectrum is shown with red open diamonds, while the solution is indicated by blue crosses. The reference synthetic spectra grid is colour-coded according to [$\alpha$/Fe] value.}
\label{Fig:GAUGUIN_fit}
\end{figure}

Once the observed spectrum is normalised, a new specific-reference synthetic spectra grid is interpolated at the input atmospheric parameters in order to measure the abundance from the analysed spectral line. This grid now includes a large range of the element abundance dimension (A$_{X}$). For the $\alpha$-elements abundance determination, the grid covers different [$\alpha$/Fe] values. A minimum quadratic distance is then calculated between the reference grid and the observed spectrum\footnote{Calculated over the wavelength domain, centred on the line, where GAUGUIN derives the abundance: \(\chi^2 = \sum_{i=1}^{N} \big[O\big(i\big) - S\big(i\big)\big]^2\), where O and S are the observed and the synthetic spectrum, respectively.}, providing a first guess of the abundance estimate (A$_{0}$). Then, this first guess is optimised via a Gauss-Newton algorithm, carrying out iterations through linearisation around the new solutions. The algorithm stops when the relative difference between two consecutive iterations is less than $\Delta$A$_{X}$\footnote{$\Delta$A$_{X}$ = 0.20 dex in the AMBRE grid}/100. Figure~\ref{Fig:GAUGUIN_fit} illustrates, for the observed solar spectrum, the fit carried out by the automated code GAUGUIN for the Mg line 5711.09 $\AA$. 

\subsection{Synthetic spectra grid} \label{GES}
A high-resolution optical synthetic grid (4200-6900\AA; R$\sim$300000, 18452 spectra) of non-rotating FGKM type spectra was used as a reference for the GAUGUIN procedure. One-dimensional LTE MARCS atmosphere models \citep{MARCS} and the spectrum synthesis code TURBOSPECTRUM for radiative transfer \citep{turbospectrum} were adopted, together with the solar chemical abundances of \citet{grevesse2007}. The covered atmospheric parameter ranges are: 3750 $\leq$ T$_{eff}$ $\leq$ 8000 K (in steps of 250 K), 0.0 $\leq$ log(g) $\leq$ 5.5 cm s$^{-2}$ (in steps of 0.5 cm s$^{-2}$), -3.5 $\leq$ [M/H] $\leq$ +1 dex (in steps of 0.25 dex), whereas the variation in [$\alpha$/Fe] is -0.4 $\leq$ [$\alpha$/Fe] $\leq$ 0.6 dex (for [M/H] $\geq$ 0.0 dex), -0.2 $\leq$ [$\alpha$/Fe] $\leq$ 0.6 dex (for -0.5 $\leq$ [M/H] < 0.0 dex), and 0.0 $\leq$ [$\alpha$/Fe] $\leq$ 0.6 (for [M/H] < -0.5 dex), with steps of 0.2 dex. \par

This grid was computed in a similar way to the original AMBRE grid \citep{patrick2012}, which was adopted for the parameter estimation. It does however contain some more recent specificities. First, we adopted the Gaia-ESO Survey atomic and molecular line lists \citep[][2019; submitted]{heiter2015}. Next, we considered more realistic values of the microturbulent velocity for the spectra computation by adopting a polynomial relation between V$_{mic}$ and the main atmospheric parameters (M. Bergemann, private communication). Finally, we always considered perfectly consistent [$\alpha$/Fe] enrichments between the selected MARCS models and the calculated emerging spectra. To be in agreement with the observational spectra data set, we reduced the resolution of the synthetic spectra to the observed resolving power (R = 115000) by convolving with a Gaussian kernel.

\subsection{Selected Mg I lines} \label{lines}

The abundance analysis was performed using nine magnesium spectral lines in the optical range shown in Table \ref{table:lines}, adopting the atomic data of \citet{heiter2015}. \par

\begin{table}[h]
\centering
\begin{tabular}{ccccc}
\hline
\hline
\multicolumn{1}{c}{\textbf{Mg I (\AA):}} \vspace{0.05cm} \\
\hline
\multicolumn{5}{c}{ \emph{Non-saturated lines:}} \\
4730.04 & 5711.09 & 6318.7 & 6319.24 & 6319.49 \vspace{0.1cm} \\
\multicolumn{5}{c}{ \emph{Saturated lines:}} \\
5167.3 & 5172.7 & 5183.6 & 5528.4 \\
\hline
\hline
\end{tabular}
\vspace{0.08cm}
\caption{Optical magnesium lines selected in the present analysis.}
\label{table:lines}
\end{table}

\begin{figure}
\centering
\includegraphics [height=65mm, width=0.48\textwidth] {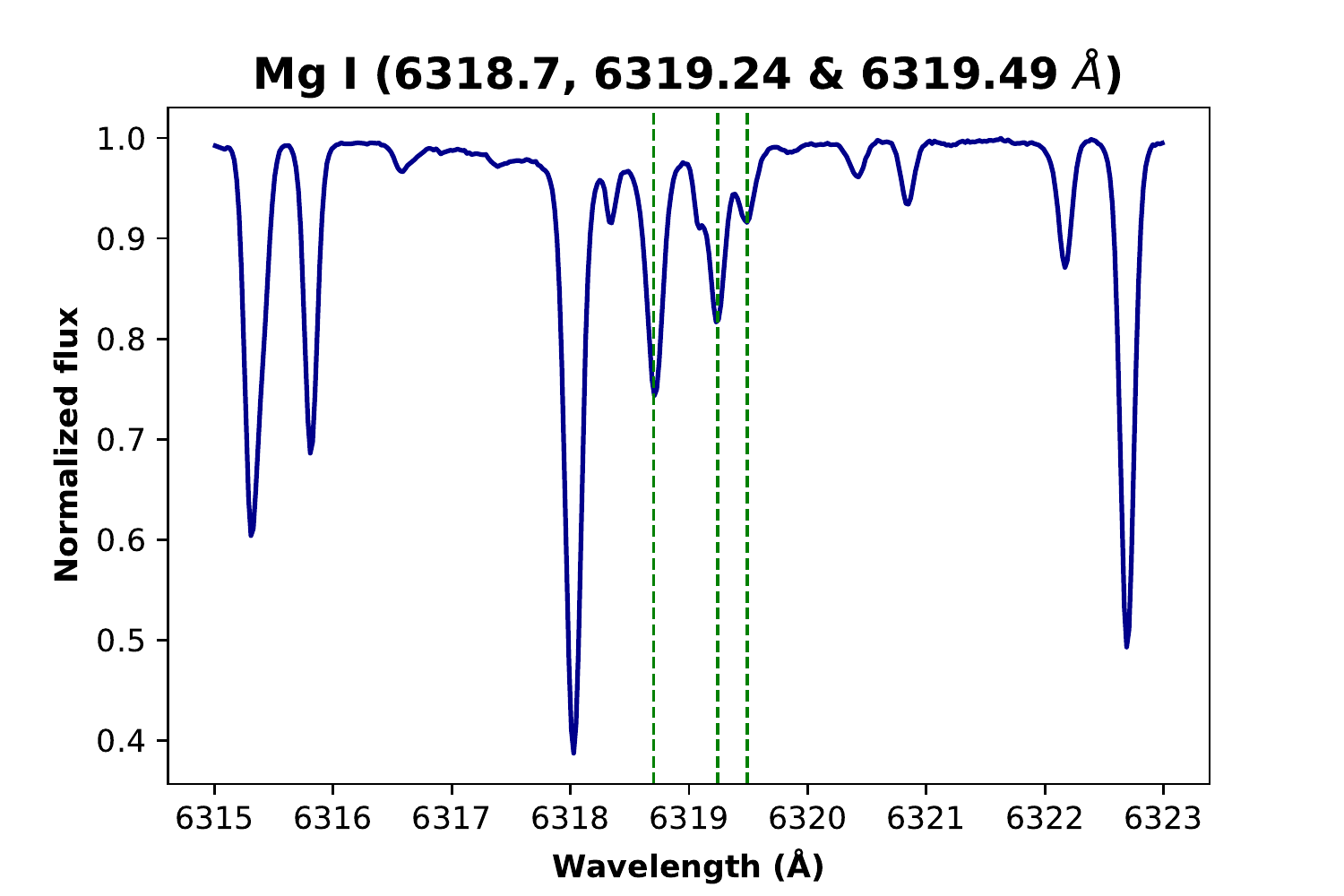}
\includegraphics [height=65mm, width=0.48\textwidth] {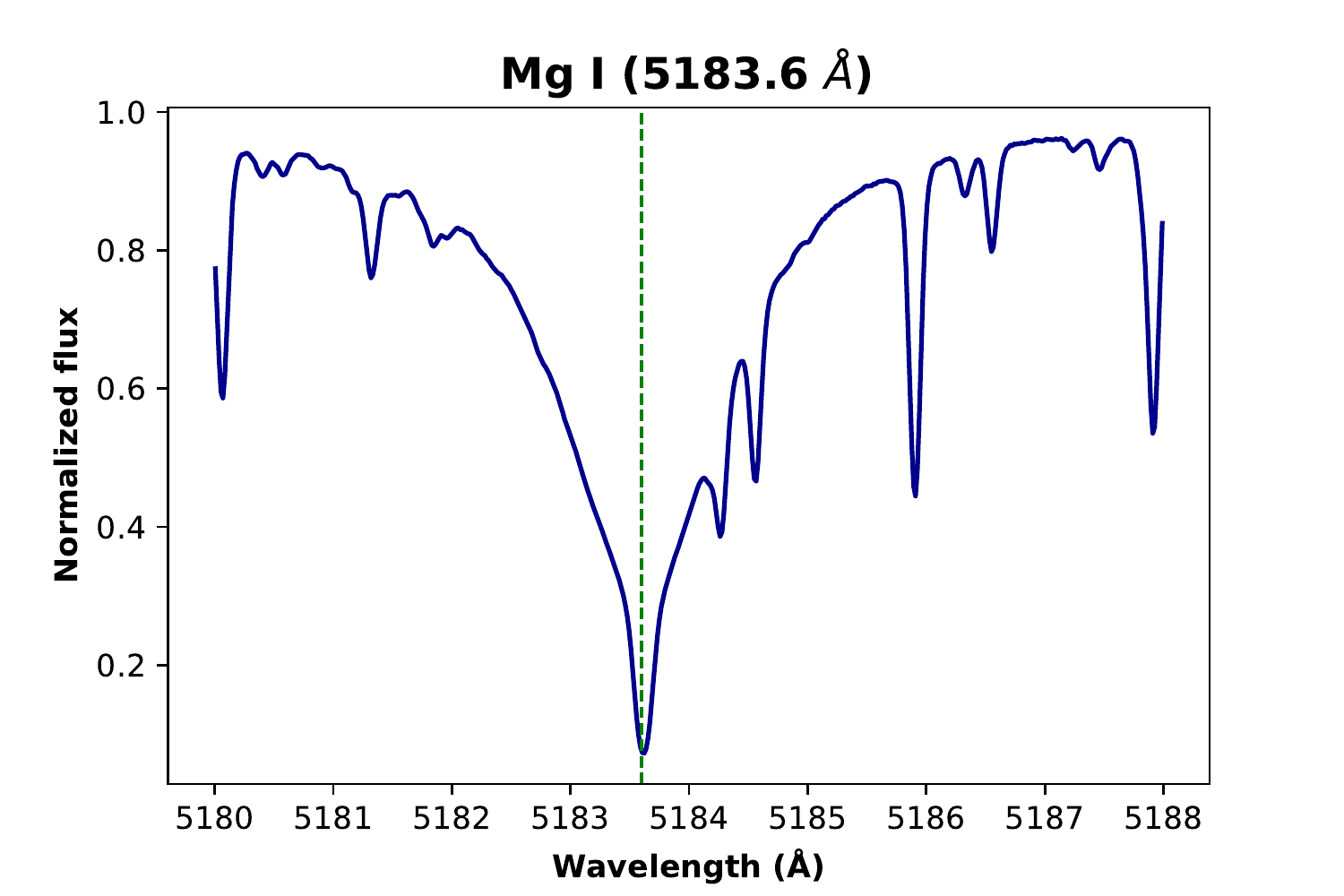}
\caption{Non-saturated triplet lines around 6319 $\AA$ (top) and the strong saturated line 5183.6 $\AA$ (bottom), identified by green dashed vertical lines, in the normalised observed solar spectrum.}
\label{Fig:sun_spectra}
\end{figure}

We performed an in-depth analysis of each line separately in order to test their reliability at different metallicity regimes. For a solar-type star, the selected lines could be classified in two categories: non-saturated (4730.04, 5711.09, and triplet: 6318.7, 6319.24, 6319.49 \AA) and strong saturated lines (Mg Ib triplet: 5167.3, 5172.7 \& 5183.6, and the line 5528.4 \AA). Both cases are illustrated in Fig.~\ref{Fig:sun_spectra}. The number of pixels available for strong lines are approximately five times higher than for non-saturated lines. We only considered non-saturated lines for spectra with FWHM$_{CCF}$ $\leq$ 7 km s$^{-1}$ to avoid possible uncertainties from line-broadening (see Appendix \ref{rotation} for further details). \par

The selected lines in Table \ref{table:lines} have been widely used in the literature to determine both [Mg/Fe] and [$\alpha$/Fe] abundances. \citet{bergemann2014,bergemann2017} analysed different approximations for radiative transfer and spectral line formation in model atmospheres, focused on their effect on Mg abundance determination using lines in the optical and infrared, among which there are four lines used in our analysis (5172, 5183, 5528, and 5711\AA). They find no significant differences between 1D LTE and 1D NLTE abundances, and for the lines in common with ours, they present a quite robust behaviour with respect to the full 3D NLTE calculations in cool FGK stars. Small NLTE effects on Mg I line formation were also found by \citet{zhao2016} and \citet{alexeeva2018}. In conclusion, 1D LTE Mg abundances are accurate enough for our selected sample and computationally cheaper than applying NLTE corrections. The results and discussions presented in this paper are therefore based on 1D LTE Mg abundances. \par

The methodology to calculate the final stellar [Mg/Fe] abundance from all the Mg lines information is described as follows. For a given spectrum, a weighted average of the individual lines results was calculated following \citet{vardan2016}, where the distance from the median abundance was considered as a weight. This method allows us to avoid the combined random uncertainties of the different lines, minimising the error when more lines are considered. Next, as at least four spectra were available for each star in the sample, the final [Mg/Fe] abundance of each object was calculated from the median value of the repeats.

%__________________________________________________________________

\section{Optimising the spectral normalisation for different stellar types}\label{normalization}

\begin{figure*}
\centering
\includegraphics [height=70mm, width=0.45\textwidth] {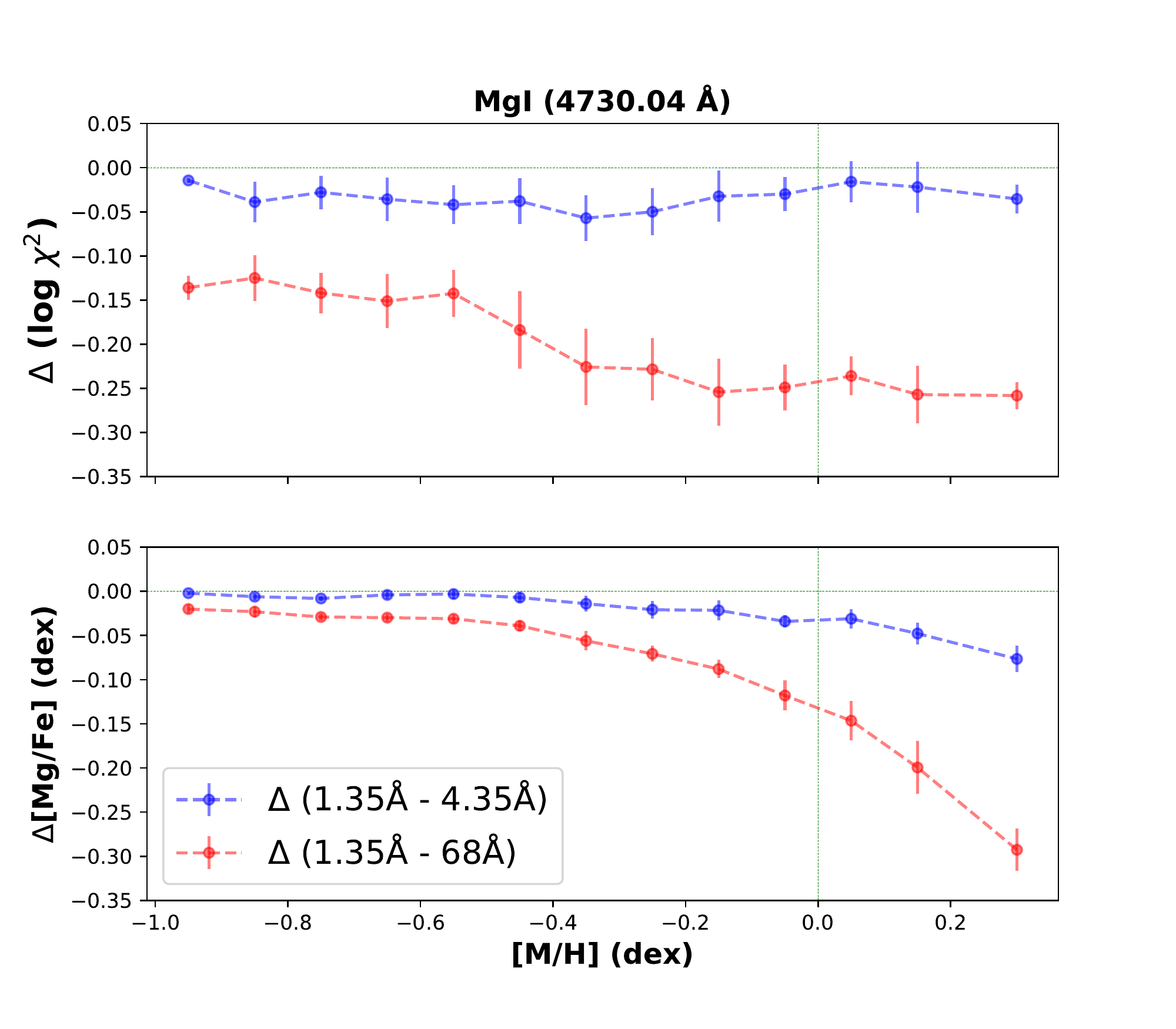}
\includegraphics [height=70mm, width=0.5\textwidth] {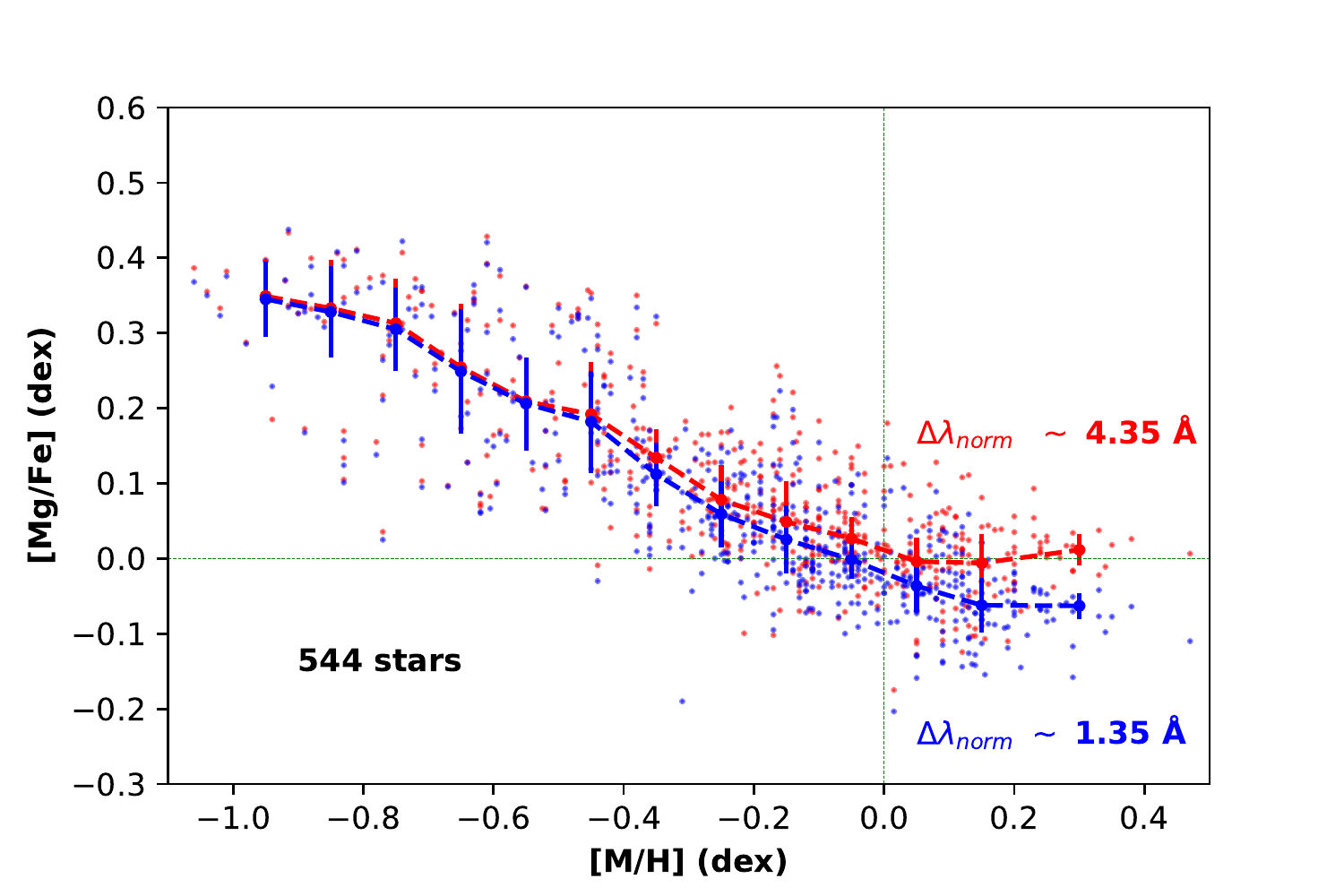}
\caption{Analysis of the non-saturated line 4730.04 $\AA$. \textbf{Left:} comparison, averaged in metallicity bins of 0.1 dex, of the line $\chi^2$ fitting in logarithmic scale (top) and the derived abundance (bottom) values from different local normalisation intervals ($\Delta\lambda_{norm}$ = 1.35, 4.35, 68 $\AA$; taking the shortest interval as a reference, which corresponds to around four times the FWHM of the line in a solar-type star). \textbf{Right:} stellar abundance ratios [Mg/Fe] vs. [M/H] for the local normalisation window $\Delta\lambda_{norm}$ = 1.35$\AA$ (blue points) and $\Delta\lambda_{norm}$ = 4.35$\AA$ (red points). Their respective behaviours were calculated by the mean [Mg/Fe] abundance value per metallicity bin. The reduced number of stars is due to the cut in FWHM$_{CCF}$ for the non-saturated lines.}
\label{Fig:norm_4730}
\end{figure*}

In large spectroscopic stellar surveys, an automatic adjustment of the continuum is performed over the observed spectrum, generally via few iterations, searching for possible line-free regions \citep{SME, sousa2007, GarciaPerez2016}. Most of the spectral analysis pipelines for determining chemical abundances and stellar atmospheric parameters carry out the same normalisation procedure for all stellar types, applying a constant continuum interval around the considered spectral feature.  \citet{sarunas2014} used a spectral fitting method to correct the local continuum in regions of $\pm$5$\AA$ and $\pm$15$\AA$ for weak and strong lines, respectively. \citet{vardan2012} applied similar normalisation intervals on the measurement of equivalent widths (private communication). The width of the constant normalisation window is assumed to find continuum information around the analysed line for any stellar type. As a consequence, the methodology is not optimised to the difficulty of identifying the continuum, which depends on the spectral type. Cool metal-rich stars\footnote{We use the following nomenclature: \par [M/H] $\lesssim$ - 0.2 dex (metal-poor); T$_{eff}$ < 5400 K (cool)  \par  [M/H] > - 0.2 dex (metal-rich);  T$_{eff}$ $\gtrsim$ 5400 K (hot)} are a particularly difficult case due to the presence of blended and molecular lines. For instance, in the case of the APOGEE survey, \citet{holtzman2015} remark how challenging it is to identify a true continuum in the observed spectra of these stellar types, leading to a 'pseudo-continuum' normalisation. Similarly, after an analysis of systematic errors using six different methods, \citet{jofre2017} concluded that the definition of continuum may be responsible for the largest fraction of the uncertainty in abundance estimations.

The automated abundance estimation code GAUGUIN is not an exception on the continuum placement performance. As described in Sect \ref{GAUGUIN}, it carries out an iterative procedure over a local window around the analysed line. For that reason, we studied the normalisation influence on the derived abundances applying (for each Mg I line; see Table \ref{table:lines}) different local continuum intervals (from narrow, $\Delta\lambda_{norm} \sim$ 1\AA, to very large ranges, $\Delta\lambda_{norm} \sim$ 70\AA. See Appendix \ref{INTERVALS}). For this purpose, we evaluated the quality of the resulting normalisation, using a goodness of fit ($\chi^2$) between the interpolated synthetic spectrum (with the corresponding atmospheric parameters of the star) and the normalised observed one. This was performed over the abundance estimation window, as it is constant for each line (c.f. blue lines in Fig.~\ref{Fig:intervals_GAUGUIN} and the corresponding fit in Fig.~\ref{Fig:GAUGUIN_fit}). \par

As described in detail hereafter, our analysis reveals that the width of the local normalisation interval can have an important impact on the derived abundances. In fact, the optimal width of the normalisation window depends clearly on the stellar type (T$_{eff}$, log g, [M/H]). As a consequence, if a constant wavelength interval is chosen, independently of the stellar parameters, different biases appear depending on the effective temperature, the surface gravity, and the global metallicity, especially in the metal-rich regime ([M/H] $\geq$ 0 dex), blurring the chemical features of the studied population. 

As expected, the environment and the intensity of the spectral line drastically influence the selection of the appropriate normalisation interval where the continuum placement should be defined for each case. In the following, we summarise the results of our study for the two characteristic cases mentioned in Table \ref{table:lines} and illustrated in Fig.~\ref{Fig:sun_spectra} for the normalised observed solar spectrum.

\subsection{Non-saturated lines}\label{weak_optimization}

Figure~\ref{Fig:norm_4730} shows, for the non-saturated line 4730.04 $\AA$, the difference in the line $\chi^2$ fitting in logarithmic scale (top-left panel) and the corresponding derived abundance values (bottom-left panel) between different local normalisation intervals. The resulting comparison reveals a more precise fit applying the narrowest interval ($\Delta\chi^2$< 0) for all the metallicities. We find that this improvement of the fit has a larger impact on the derived abundances for the metal-rich stars, leading to lower [Mg/Fe] abundances at supersolar metallicities, with differences as high as $\sim$0.3 dex at [M/H]= +0.2 dex in comparison with the largest wavelength domain. The right panel of Fig.~\ref{Fig:norm_4730}, showing the [Mg/Fe] vs. [M/H] plane, highlights the influence of the normalisation procedure on the behaviour of the $\alpha$-elements in the metal-rich regime of the disc. \par

\begin{figure}
\centering
\includegraphics [height=65mm, width=0.4\textwidth] {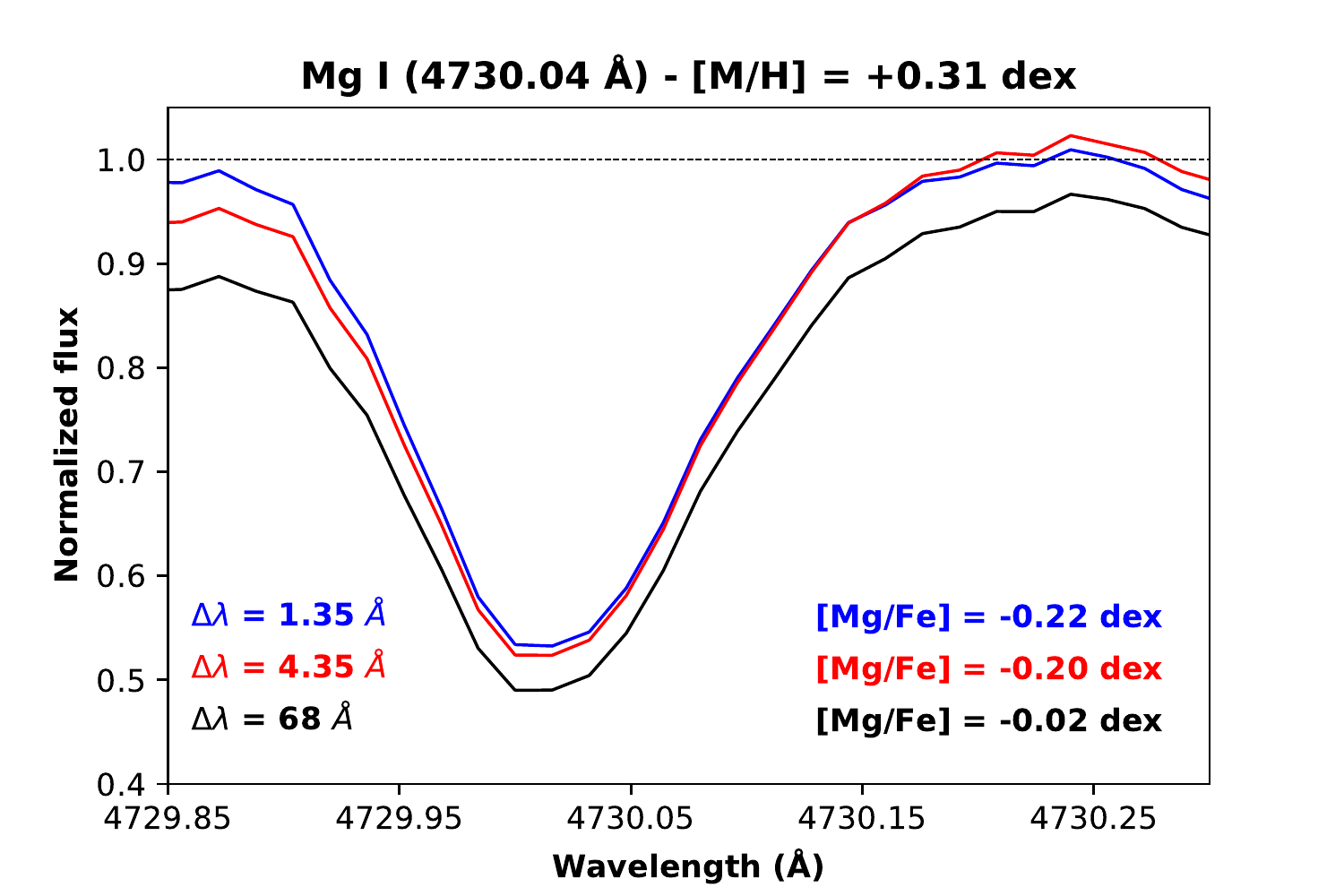}
\caption{Normalised observed spectrum of a particular metal-rich star ([M/H]= + 0.31 dex) in the abundance estimation window of the non-saturated line 4730.04 $\AA$. Different line profile depending on the applied local continuum intervals ($\Delta\lambda_{norm}$ = 1.35 (blue), 4.35 (red), 68 $\AA$ (black)), along with the derived [Mg/Fe] abundance for each case.}
\label{Fig:norm_4730_Flux}
\end{figure}

Figure~\ref{Fig:norm_4730_Flux} illustrates the effect of different normalisation windows in the flux of the Mg line at 4730.04 $\AA$, for a particular metal-rich star ([M/H]= + 0.31 dex). The observed flux variations are responsible for the abundance differences observed in Fig.~\ref{Fig:norm_4730}. The use of a larger normalisation interval can drop artificially the observed flux, leading to higher abundance estimates. More difficult pseudo-continuum placements due to contiguous absorption lines can explain this continuum drop, and the poorer goodness-of-fit values observed in Fig.~\ref{Fig:norm_4730}. \par

As for the 4730.04 $\AA$ line, the weak triplet lines around 6319$\AA$ (left panel in Fig.~\ref{Fig:sun_spectra}) are better fitted by applying a narrow normalisation interval. For the Mg line 5711.09 $\AA$ (see Fig.~\ref{Fig:intervals_GAUGUIN}), which is stronger than the other non-saturated lines, the optimal local normalisation window is larger for cool metal-rich stars, for which the line is more intense although still not saturated. \par

In conclusion, for non-saturated lines, there is generally enough continuum information around and close to the line. As a consequence, it is convenient to optimise the normalisation interval, close to the considered spectral feature.

\subsection{Strong saturated lines}\label{intense_optimization}

For strong saturated lines like the Mg Ib triplet (5167.3, 5172.7, and 5183.6 $\AA$) and the line 5528.4 $\AA$, no pixels are available close to the continuum level for most of the stellar types (see bottom panel in Fig.~\ref{Fig:sun_spectra}) in the analysed region, and a pseudo-continuum normalisation has to be performed for the automatic fit, including part of the line wings. \par

\begin{figure}
\centering
\includegraphics[height=115mm, width=0.42\textwidth]{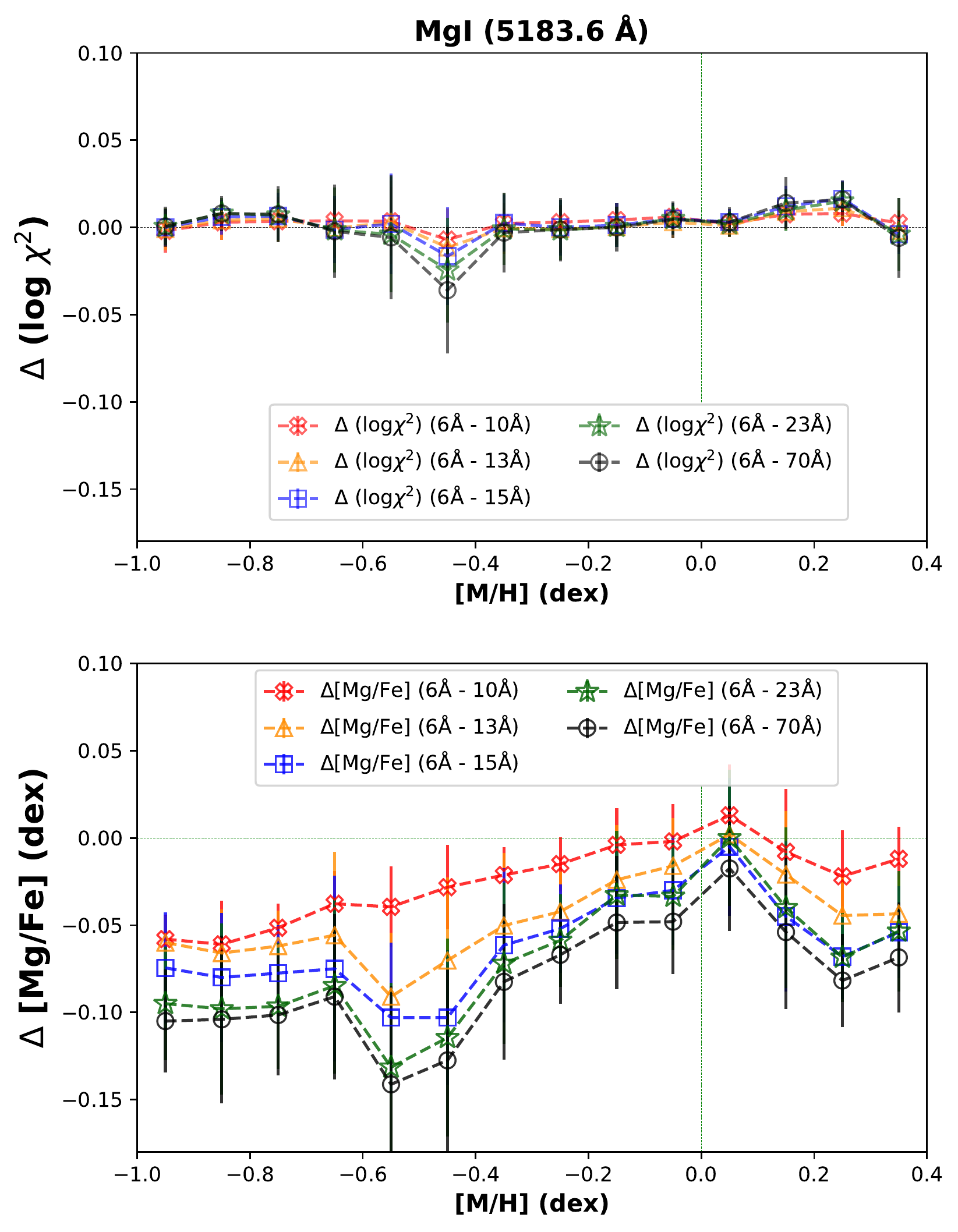}
\caption{Analysis of strong saturated line 5183.6 $\AA$. \textbf{Top:} comparison, averaged in metallicity bins of 0.1 dex, of the line $\chi^2$ fitting values (in logarithmic scale) for different local normalisation intervals ($\Delta\lambda_{norm}\sim$6$\AA$, 10$\AA$, 13$\AA$, 15$\AA$, 23$\AA$, 70$\AA$; taking the shortest interval as a reference, which corresponds to two times the FWHM of the line in a solar-type star, approximately). \textbf{Bottom:} same analysis comparing the derived abundance values.} 
\label{Fig:Mg3_chi2}
\end{figure}

However, two main difficulties affect the procedure. On the one hand, due to the line saturation, only the wings are sensitive to the abundance. As a consequence, an important degeneracy between the continuum placement and the derived [Mg/Fe] abundance appears. In other words, large changes in the continuum placement, like those induced by the use of different normalisation windows, can be compensated by a change in the abundance without degrading the line fitting quality. Figure~\ref{Fig:Mg3_chi2} illustrates,\, for the different local normalisation intervals analysed around the 5183.6$\AA$ line, the comparison of the line $\chi^2$ fitting values (top panel), and the corresponding derived abundance values (bottom panel). A negligible difference in the goodness of fit from the studied continuum intervals is observed, although, as shown below, it corresponds to notorious differences in the abundance estimations. Therefore, the $\chi^2$ quality criterion, reliable to carry out an appropriate normalisation interval selection for the non-saturated lines (Sect. \ref{weak_optimization}), is usually not discriminating enough in saturated lines. \par

\begin{figure*}
\centering
\includegraphics[height=110mm, width=\textwidth]{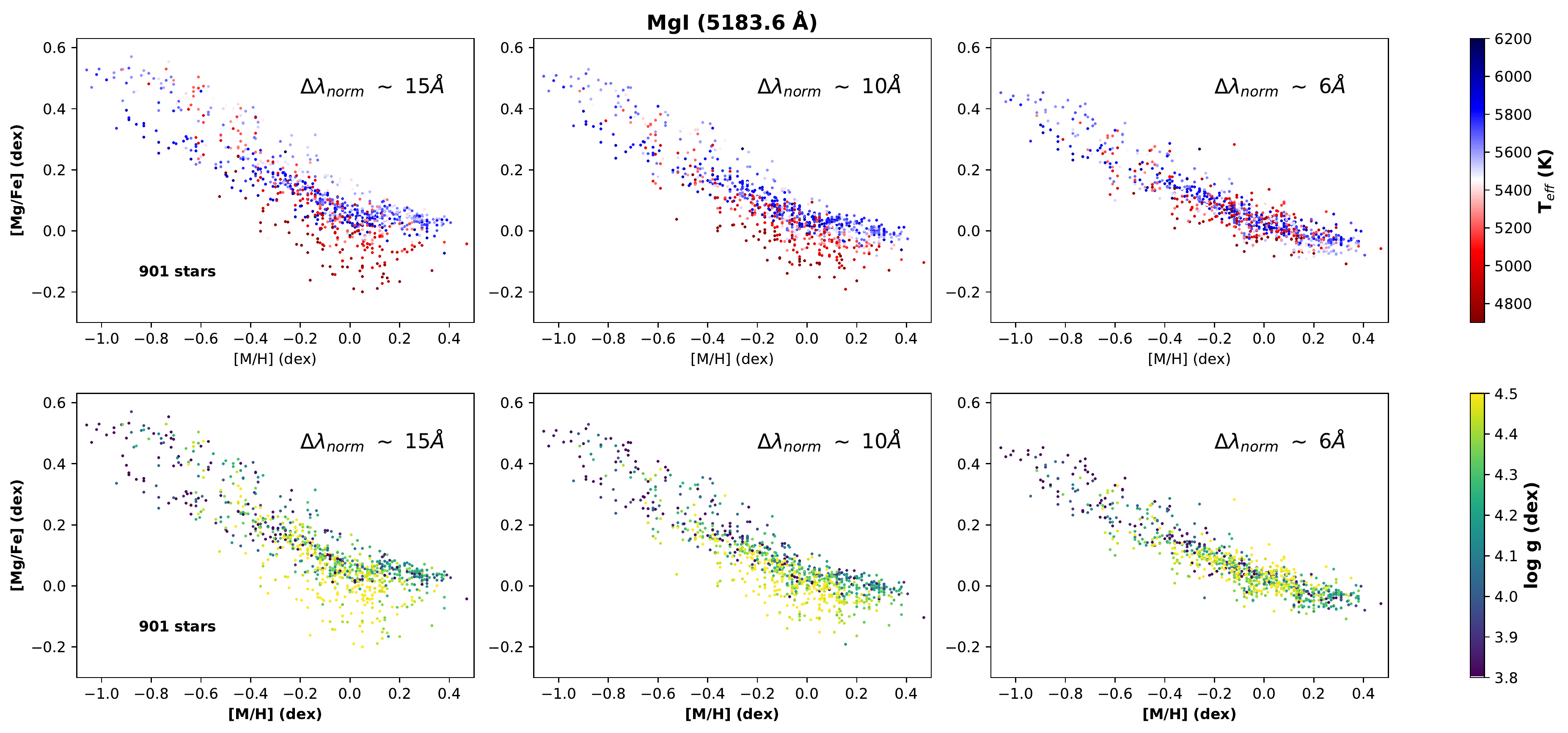}
\caption{Comparison of stellar abundance ratios [Mg/Fe] vs. [M/H] derived for the strong saturated line 5183.6 $\AA$, colour-coded by stellar effective temperature (top row) and surface gravity (bottom row), after carrying out the continuum placement in three different local normalisation intervals around the line. \textbf{Left:} $\Delta\lambda_{norm}\sim$ 15$\AA$, typical local continuum interval applied in the literature for strong lines. \textbf{Middle:} $\Delta\lambda_{norm}\sim$ 10$\AA$. \textbf{Right:} $\Delta\lambda_{norm}\sim$ 6$\AA$, narrow normalisation interval ($\sim$ twice the FWHM of the line in solar-type stars).}
\label{Fig:Mg3_teff_logg}
\end{figure*}

On the other hand, the larger the local normalisation window, the larger the dependencies of the abundance results on the parameters and, as a consequence, the larger the dispersion on the [Mg/Fe] abundance with respect to [M/H]. This is illustrated in Fig.~\ref{Fig:Mg3_teff_logg}, where the resulting [Mg/Fe] vs. [M/H] abundances are shown for the same line and most representative normalisation intervals of Fig.~\ref{Fig:Mg3_chi2}, colour-coded with the star's effective temperature and surface gravity. Clearly, the results obtained with the largest normalisation window (left panels) have a significant T$_{eff}$ dependence and even a log g dependence, inducing a higher dispersion. Those effects are alleviated when the normalisation window is narrowed to 10$\AA$ (middle panels), and they practically disappear for the narrowest window of 6$\AA$ (right panels). In addition, broader windows tend to have lower [Mg/Fe] values for cooler and higher gravity stars. These trends do not disappear even if an iterative procedure involving local normalisation and abundance estimation  is implemented. In addition, the parameter dependence is also observed when larger normalisation intervals are explored (going beyond the MgI triplet wings, up to $\Delta\lambda_{norm}\sim$ 70$\AA$; 5140 - 5210$\AA$). In the following, we analyse in detail the reason for this observed pattern of strong saturated lines. \par

\begin{figure*}
\centering
\includegraphics[height=72mm, width=\textwidth]{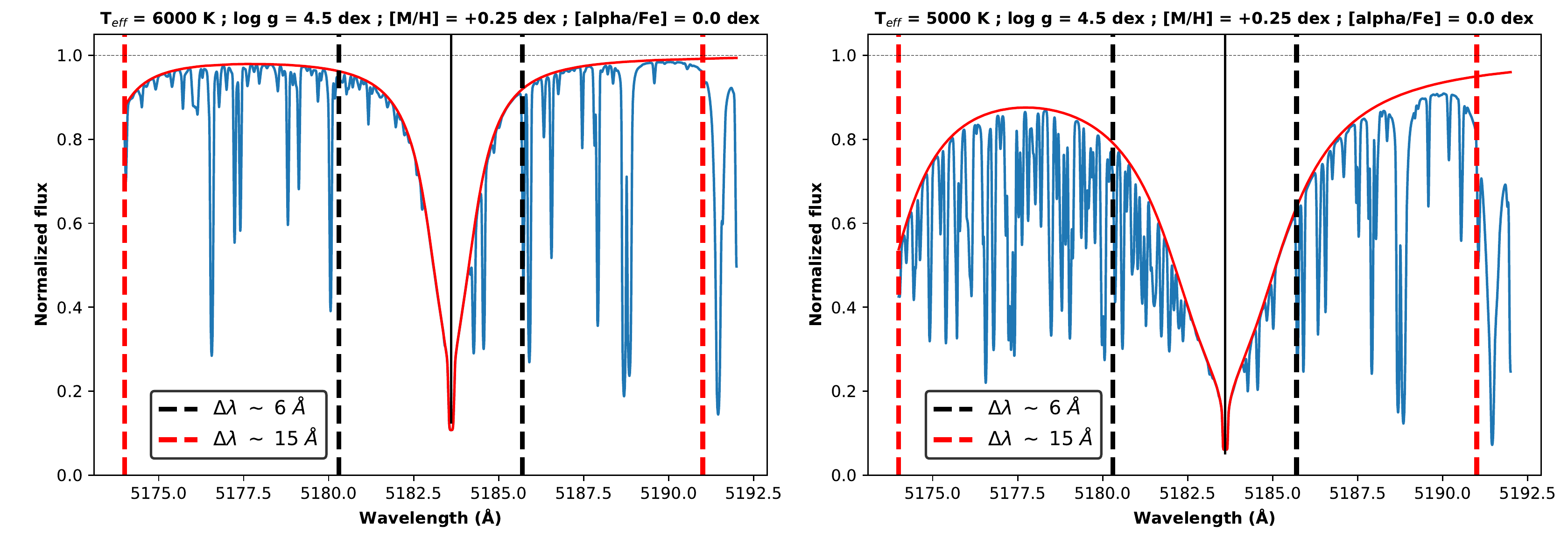}
\caption{Comparison of synthetic spectra around the saturated line 5183.6 $\AA$ (black solid vertical line), considering all the absorption lines (blue spectra) and only the Mg line absorption (red spectra). The black and red dashed vertical lines present the limits of the 6$\AA$ and 15$\AA$ normalisation windows, respectively. Left: hot star (T$_{eff}$ = 6000 K). Right: cool star (T$_{eff}$ = 5000 K). The other atmospheric parameters are constant (log g = 4.5 dex, [M/H] = + 0.25 dex, [$\alpha$/Fe] = 0.0 dex). }
\label{Fig:Mg3_synthetic}
\end{figure*}

\begin{figure*}[h]
\centering
\includegraphics[height=60mm, width=0.48\textwidth]{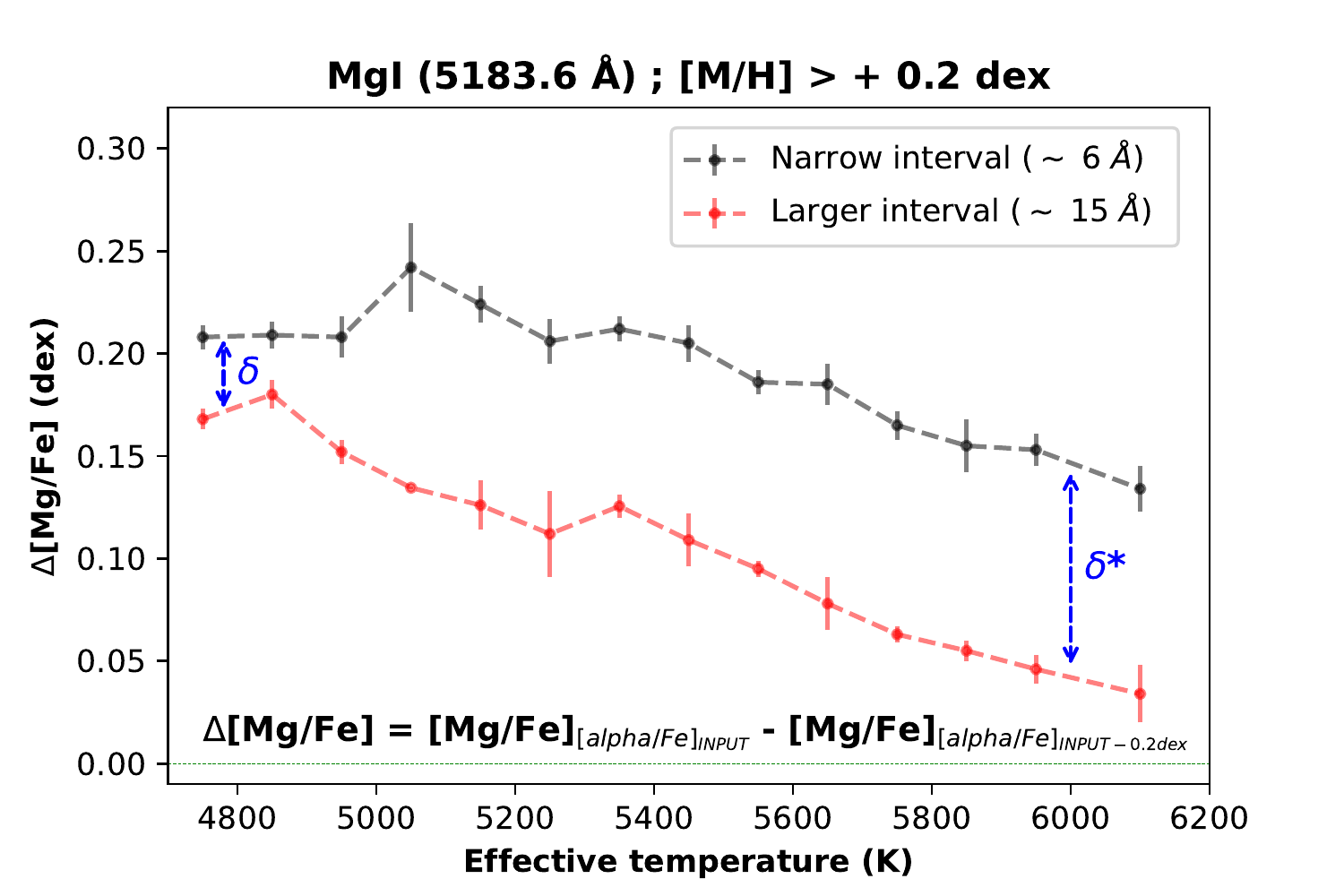}
\includegraphics[height=60mm, width=0.49\textwidth]{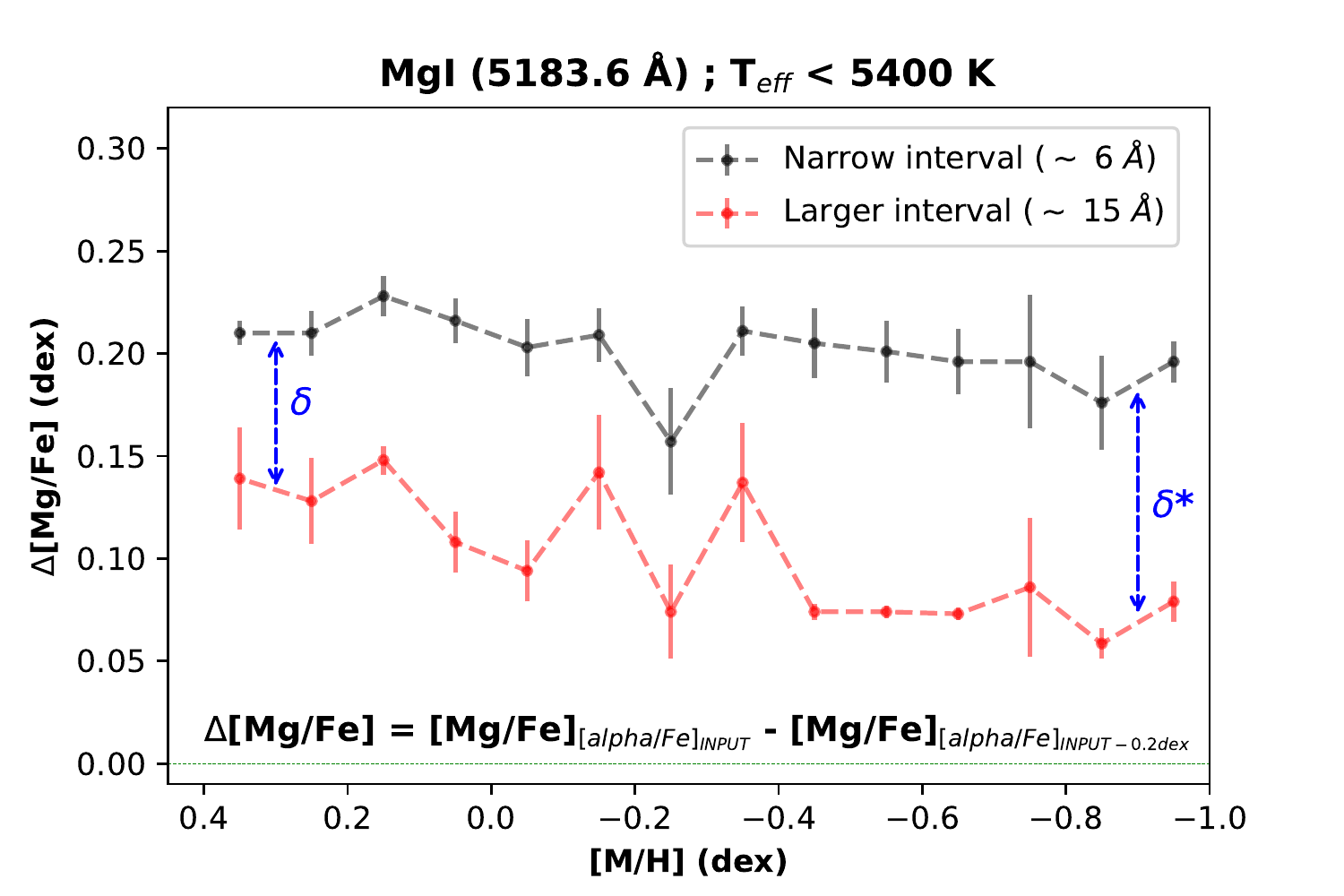}
\caption{Analysis of saturated line 5183.6 $\AA$. Comparison of the derived [Mg/Fe] abundances per spectrum after introducing a bias in the input [$\alpha$/Fe] value ([$\alpha$/Fe]$^\star$ = [$\alpha$/Fe] - 0.2 dex) to simulate a [$\alpha$/Fe]-[Mg/Fe] shift. The black and red lines correspond to the normalisation windows of 6Å and 15Å, respectively. \textbf{Left:} temperature dependence for the metal-rich sample ([M/H] $\gtrsim$ + 0.2 dex). \textbf{Right:} metallicity dependence for the cool sample (T$_{eff}$ $\lesssim$ 5400K).}
\label{Fig:Mg3_bias}
\end{figure*}

\begin{figure*}
\centering
\includegraphics[height=60mm, width=0.9\textwidth]{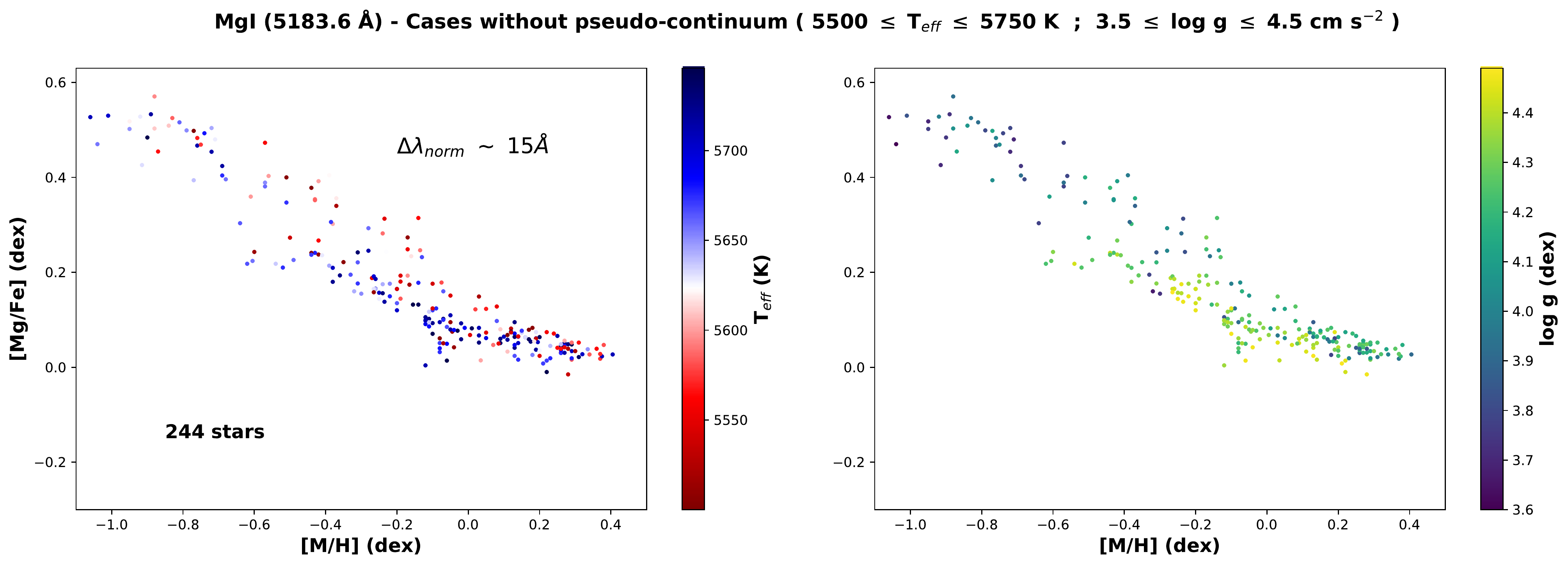}
\caption{Stellar abundance ratios [Mg/Fe] vs. [M/H], including only cases without pseudo-continuum, for the large normalisation window of 15$\AA$ around the strong line 5183.6 $\AA$, colour-coded by stellar effective temperature (left panel) and surface gravity (right panel). }
\label{Fig:fiables}
\end{figure*}

To better illustrate this behaviour, Fig.~\ref{Fig:Mg3_synthetic} shows the synthetic spectra for a hot star (T$_{eff}$ = 6000 K, left panel) and a cool star (T$_{eff}$ = 5000 K, right panel), keeping the other atmospheric parameters constant to the following values: 4.5 dex in logg, + 0.25 dex for metallicity, and 0.0 dex in [$\alpha$/Fe] abundance. The blue synthetic spectra consider all the absorption sources (atomic and molecular lines), while the red ones present only the Mg lines absorption source for comparison purposes. As illustrated in the right panel, the pseudo-continuum level is essentially driven by the Mg line absorption. As a consequence, when no pixels at the continuum are available (e.g. cool and metal-rich stars), the first Mg abundance guess has a strong influence in the pseudo-continuum placement. In our procedure, the normalisation step, taking as a reference a synthetic spectrum with the star's atmospheric parameters, assumes as a first guess that [Mg/Fe] is equal to the input [$\alpha$/Fe] parameter (see Sect. \ref{Normalisation}). 
 \par

To explore the dependence of the normalisation, and therefore of the resulting [Mg/Fe] estimates, on the initial abundance guess, we artificially substracted 0.2 dex to the corresponding [$\alpha$/Fe] initial value of each star ([$\alpha$/Fe]$^\star_{input}$ = [$\alpha$/Fe]$_{input}$ - 0.2 dex) without modifying the observed spectra sample, simulating a difference between the [$\alpha$/Fe] and the [Mg/Fe] abundance. Figure~\ref{Fig:Mg3_bias} shows the difference in the resulting [Mg/Fe] estimate for two local normalisation windows of 6 and 15 $\AA$ around the saturated line 5183.6 $\AA$ (illustrated in Fig.~\ref{Fig:Mg3_synthetic}). On the left panel, the temperature dependence of the differences is plotted only for the metal-rich sample. On the right panel, the metallicity dependence of the differences is only plotted for the cool sample. This allows us to isolate the impact of the [$\alpha$/Fe]-[Mg/Fe] bias for the typical cases of spectra with pseudo-continuum. The normalisation procedure is independent on the initial [$\alpha$/Fe] when the derived [Mg/Fe] abundance has not been shifted ($\Delta$[Mg/Fe]$_{measured}$ = 0.0 dex). For the large normalisation window (red curve), we observe that the derived [Mg/Fe] abundance depends highly on the assumed parameters (temperature on the right panel and metallicity on the left one). Indeed, as the number of pixels at the real continuum used to normalise varies from nearly 100\% (hot, metal-poor stars) to 0\% (cool, metal-rich stars), the parameter dependence of the result will vary from no impact ($\Delta$[Mg/Fe]$_{measured}$ = 0) to an almost completely dependent situation ($\Delta$[Mg/Fe]$_{measured}$ = + 0.2 dex, that is the introduced bias). In the narrow normalisation window (black curve), all the stellar types have a very low number of pixels at the continuum, implying only small adjustments with respect to the initial implicit guess given by the assumed [$\alpha$/Fe]. Therefore, almost no parameter dependence is observed around a $\Delta$[Mg/Fe]$_{measured}$=+0.2 dex. \par

This result is in close agreement with the synthetic analysis showed in Fig.~\ref{Fig:Mg3_synthetic}. The induced bias in the continuum-level estimation, and therefore in the derived [Mg/Fe] abundances, is always present but constant with the stellar atmospheric parameters if the narrow normalisation interval is applied around the strong saturated lines. This is not the case for the 15$\AA$ window, for which this induced bias is partially corrected but only to an extent that depends on the stellar parameters. \par

The application of a large local continuum interval is only convenient for the stellar types with pixels reaching the continuum level. In those cases, as described before, the normalisation procedure is independent on the initial guess of the [$\alpha$/Fe] value. In our sample, this condition only occurs for stars with 5500 $\leq$ T$_{eff}$ $\leq$ 5750 K  and  3.5 $\leq$ log g $\leq$ 4.5 dex. The corresponding [Mg/Fe] vs. [M/H] for these stars, using the large normalisation window (15$\AA$), is shown in Fig.~\ref{Fig:fiables}. It can be appreciated that the thin and thick disc sequences clearly separate up to a metallicity of $\sim$ +0.1 dex, and the trends at high metallicity do not flatten. These more accurate results are in agreement with the observed trends for the narrowest (6$\AA$) normalisation window for all the stellar types (right panels of Fig.~\ref{Fig:Mg3_teff_logg}). We therefore conclude that our procedure, through narrow normalisation windows using the global [$\alpha$/Fe] as an initial abundance guess, is a reliable way of using these strong saturated lines. \par

In conclusion, the dispersion on the [Mg/Fe] abundance estimation from strong saturated lines, with respect to [M/H], is dominated by the induced bias in the continuum-level estimation for the stellar types where a pseudo-continuum evaluation around the line is performed. The application of larger normalisation windows results in a parameter dependence of the obtained abundance and a larger line-to-line dispersion, each saturated line having its own level of continuum misplacement for a given star. The amplitude of this continuum placement error is smaller applying a narrower normalisation interval, therefore improving the abundance estimation precision. The strong link of the narrow normalisation window to the initial [$\alpha$/Fe] guess through the pseudo-continuum reduces the atmospheric parameter dependence.

\subsection{Mg abundances line-to-line scatter}\label{optimal_analysis}

The previous sections allow us to conclude that: (i) for weak non-saturated lines, the optimal wavelength domain for the local continuum placement has to be evaluated using a goodness-of-fit criterion, allowing a wavelength dependence with the spectral type. Generally, narrow normalisation windows between 1 and 2 $\AA$ provide the best line fittings (around two to four times the FWHM of each line in a solar-type star); (ii) for strong saturated lines, a narrow normalisation window allows us to reduce parameter-dependent biases of the abundance estimate, improving the precision (around two to four times the FWHM of each line in a solar-type star). \par

\begin{figure}
\centering
\includegraphics[height=62mm, width=0.45\textwidth]{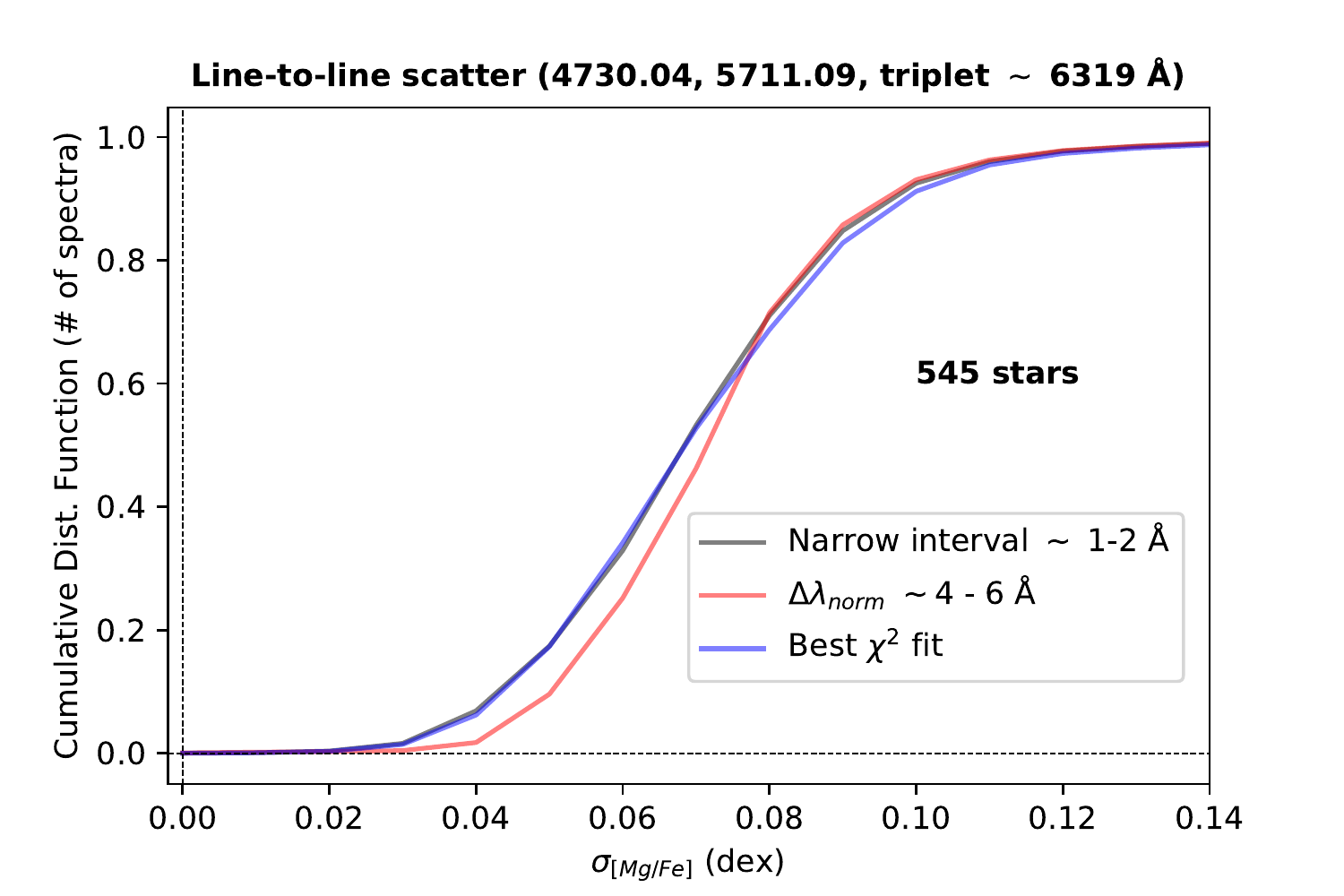}
\includegraphics[height=62mm, width=0.45\textwidth]{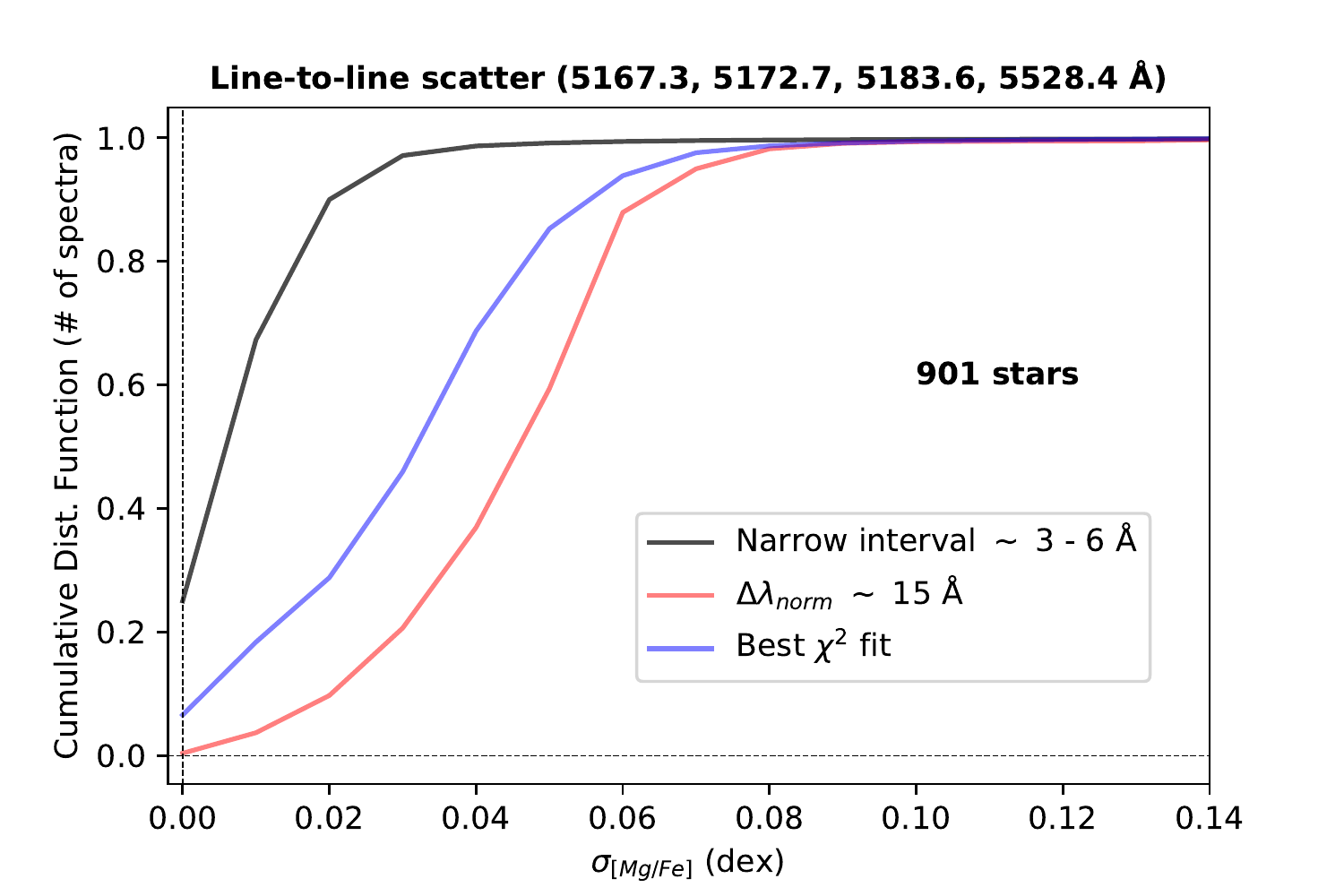}
\caption{Cumulative distribution function of the line-to-line scatter estimation of the derived [Mg/Fe] abundances per spectrum, for weak non-saturated Mg lines (top; 4730.04, 5711.09, 6318.7, 6319.24, and 6319.49$\AA$) and strong saturated ones (bottom; 5167.3, 5172.7, 5183.6, and 5528.4 $\AA$) separately. The blue curve describes the values that correspond to the normalisation interval that presents the lowest $\chi^2$ fitting value, the red one corresponds to the continuum placement performance in the typical wavelength interval used in previous works in the literature, and the black curve corresponds to a narrower local normalisation interval around each line as proposed in this work.}
\label{Fig:internal_error_weakVSintense}
\end{figure}

To evaluate the improvement in precision of the abundance results with our optimised procedure, we analysed the internal error estimation. Figure~\ref{Fig:internal_error_weakVSintense} shows the cumulative distribution of the line-to-line scatter per spectrum of the derived [Mg/Fe] abundances for non-saturated (top panel) and strong saturated lines (bottom panel) separately. For non-saturated lines, the improvement in precision is confirmed after the optimisation based on the goodness of fit (blue curve), and applying narrow normalisation windows around each line ($\Delta\lambda_{norm}$ $\sim$ 1-2 $\AA$) shows the same behaviour in terms of line-to-line scatter (black curve). The reduced number of stars is due to the cut in FWHM$_{CCF}$ (see Sect. \ref{lines}). For strong saturated lines, the application of a narrow normalisation interval ($\Delta\lambda_{norm}$ $\sim$ 3 - 6 $\AA$, two to four times their FWHM in a solar-type star) shows a remarkable improvement of the abundance estimation precision. It presents a small scatter of less than 0.03 dex in comparison with the continuum placement performance in the typical wavelength interval used in previous works (red curve) and with the consideration of the goodness of fit as the quality criterion of the resulting normalisation, despite the fact that we showed that it is not discriminating enough for these lines (see Fig.~\ref{Fig:Mg3_chi2}). In conclusion, our optimised normalisation procedure applies a goodness-of-fit criterion for weak non-saturated lines to choose the appropriate continuum interval, and a fixed narrow normalisation window for the strong saturated ones. \par

\begin{figure}
\centering
\includegraphics[height=63mm, width=0.5\textwidth]{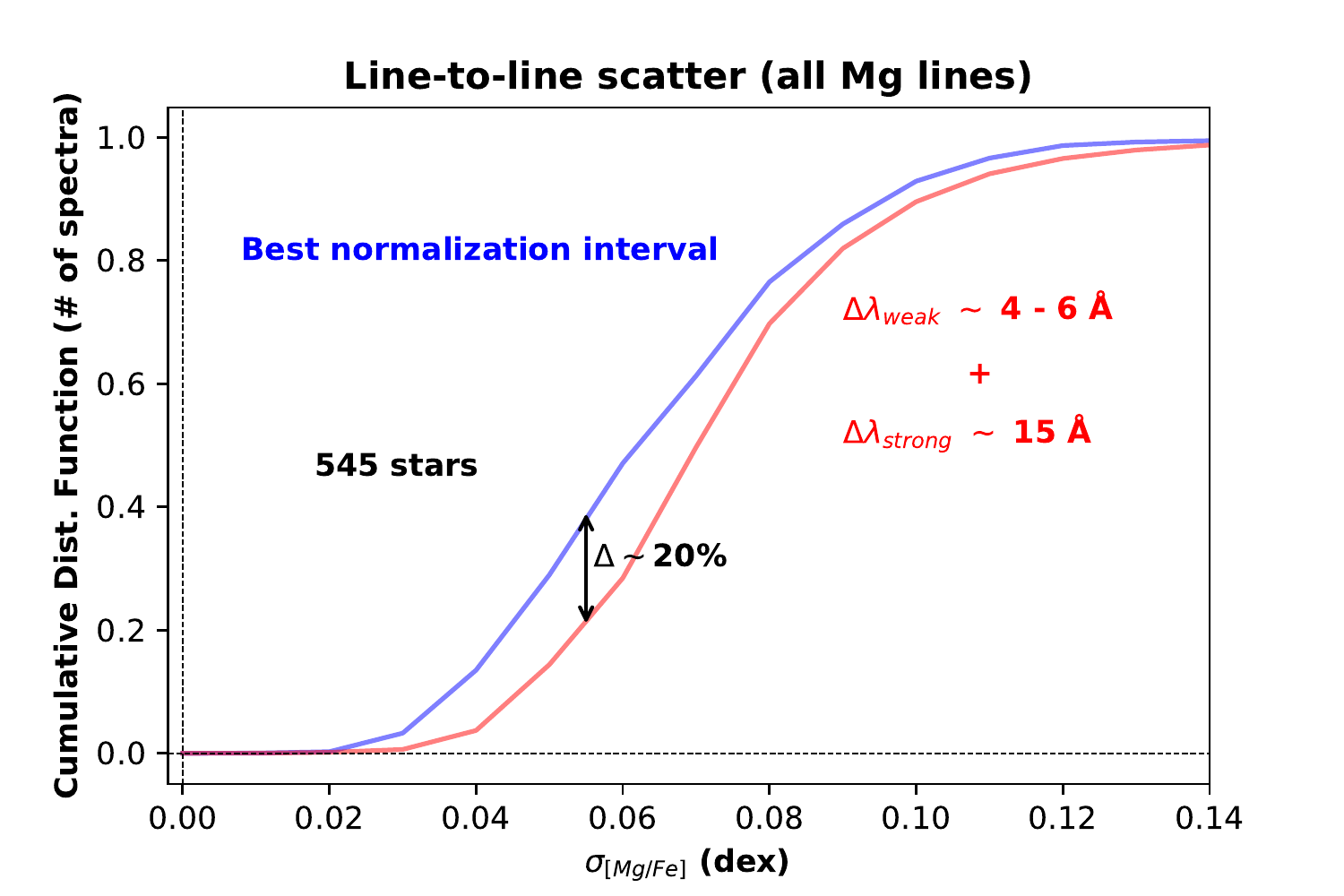}
\includegraphics[height=63mm, width=0.5\textwidth]{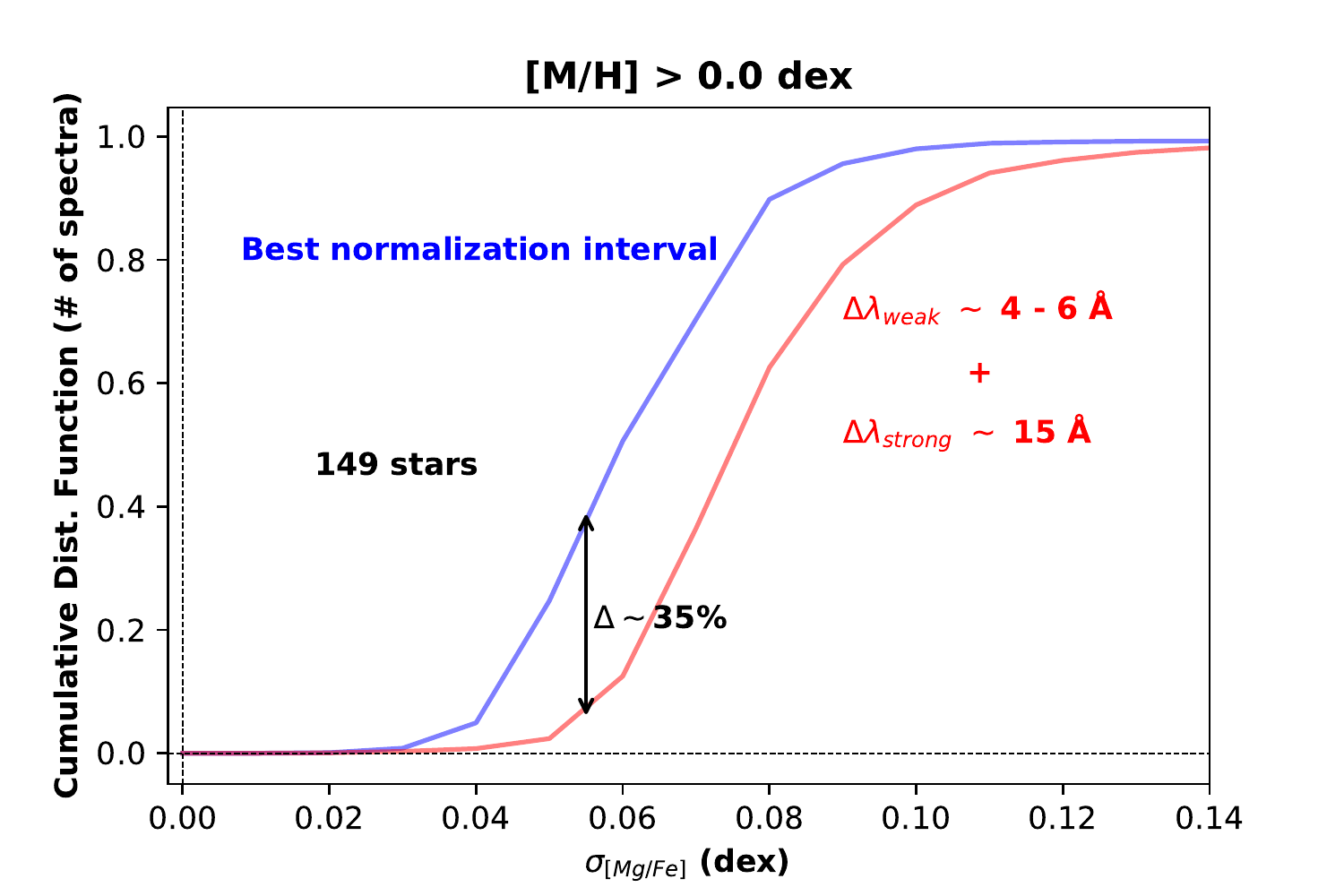}
\caption{\textbf{Top:} cumulative distribution function of the line-to-line scatter estimation of the derived [Mg/Fe] abundance ratio for the spectra sample for which all the Mg lines are taken into account.  \textbf{Bottom:} same but for the metal-rich sample ([M/H] > 0.0 dex). The red curve corresponds to the typical normalisation windows used in the literature, while the blue curve shows the results with our optimised normalisation procedure.}
\label{Fig:internal_error}
\end{figure}

A similar behaviour is observed in Fig. \ref{Fig:internal_error} when all the Mg lines of different strengths are considered. We compare results for the normalisation windows used in the literature (red) and for our optimised normalisation procedure (blue). In the top panel, we can see the analysis over the whole spectra sample for which all the Mg lines are considered. In addition, 40 \% of the sample presents a scatter smaller than 0.05 dex choosing our best value in each case, while this is only the case for 20 \% of the sample with the classical normalisation windows. This improvement in precision is even more significant if we focus the analysis on the metal-rich part ([M/H] > 0.0 dex), as shown in the bottom panel of Fig. \ref{Fig:internal_error}. In the metal-rich regime, the percentage of spectra with a line-to-line scatter lower than 0.05 dex increases from 5 \% to 40 \%, thanks to our optimised procedure with respect to the classical one. \par

In conclusion, this reduction of the line-to-line scatter supports an improvement in the abundance estimate's precision with our procedure.  In terms of accuracy, we identified four Gaia-benchmark stars (18 Sco, HD 22879, Sun, and $\tau$ Cet) from \citet{jofre2015} in our sample, finding an excellent agreement with an overall average difference of 0.01 dex. We also tested the derived abundances for those stars applying the normalisation windows used in the literature, finding an average difference within one sigma error from the optimised value. In other words, our normalisation procedure improves the abundance estimation precision, preserving its  accuracy. \par 

%---------------------------------------------------------------------------------------------------

\section{Disentangling the thin and the thick disc populations} \label{results}

\begin{figure*}
\centering
\includegraphics [height=180mm, width=0.42\textwidth] {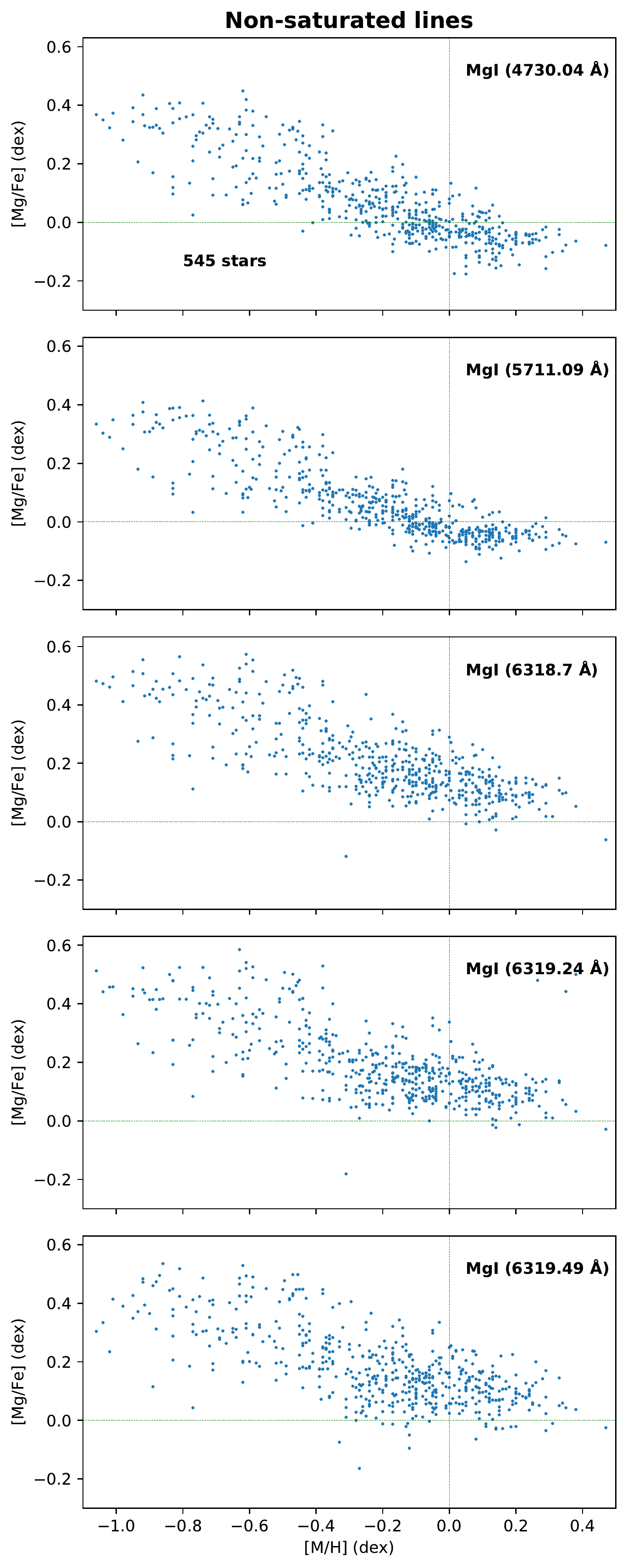}
\includegraphics [height=180mm, width=0.45\textwidth] {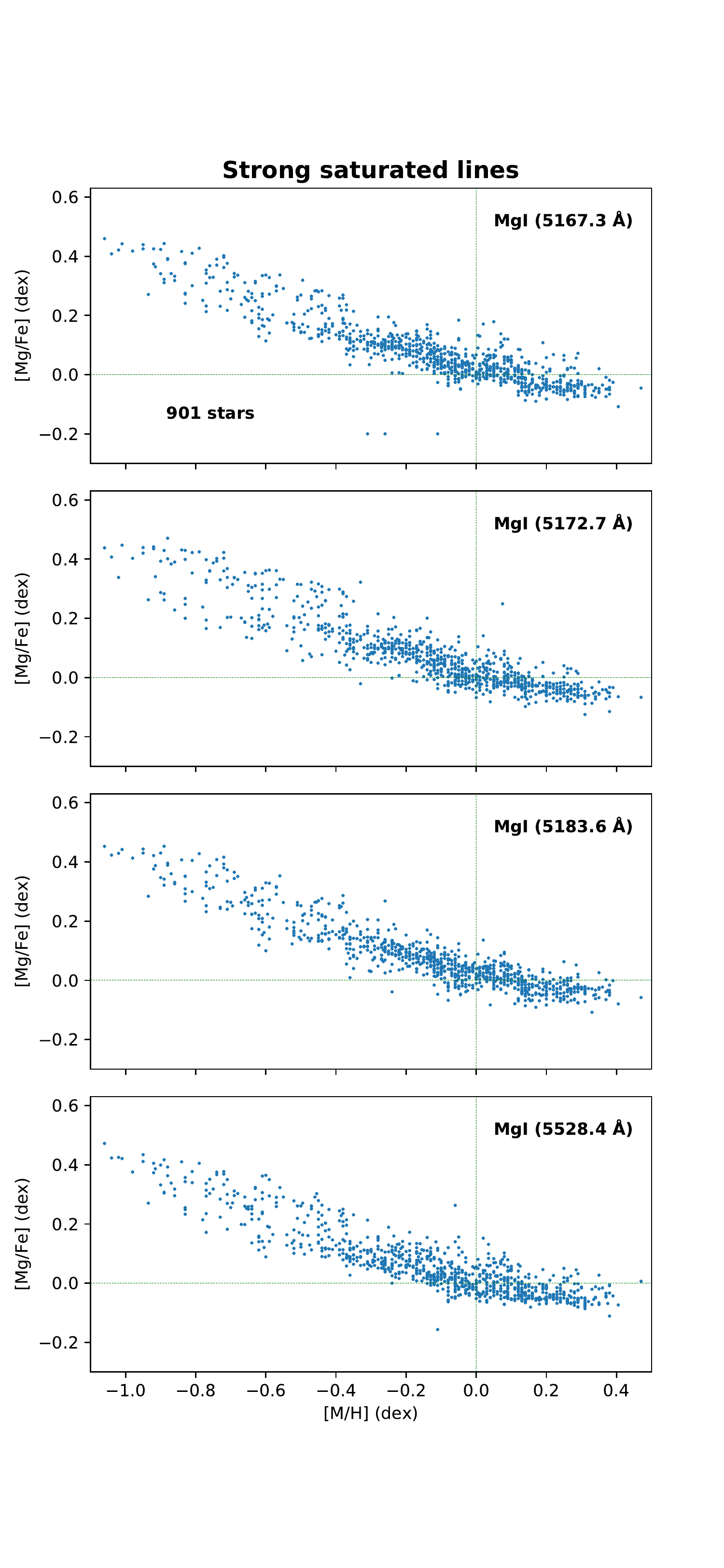}

\caption{Most precise stellar abundance ratios [Mg/Fe] vs. [M/H] following the optimal method for each Mg I spectral line separately. \textbf{Left:} non-saturated lines: 4730.04, 5711.09, 6318.7, 6319.24, and 6319.49 $\AA$. \textbf{Right:} strong saturated lines: 5167.3, 5172.7, 5183.6, and 5528.4 $\AA$ (from top to bottom).}
\label{Fig:lines_mejorchi2}
\end{figure*}

In this section, we summarise the final derived [Mg/Fe] abundances of our AMBRE:HARPS stellar sample (901 stars, see Fig. \ref{Fig:HR_clean}).

\begin{figure*}
\centering
\includegraphics [height=70mm, width=0.55\textwidth] {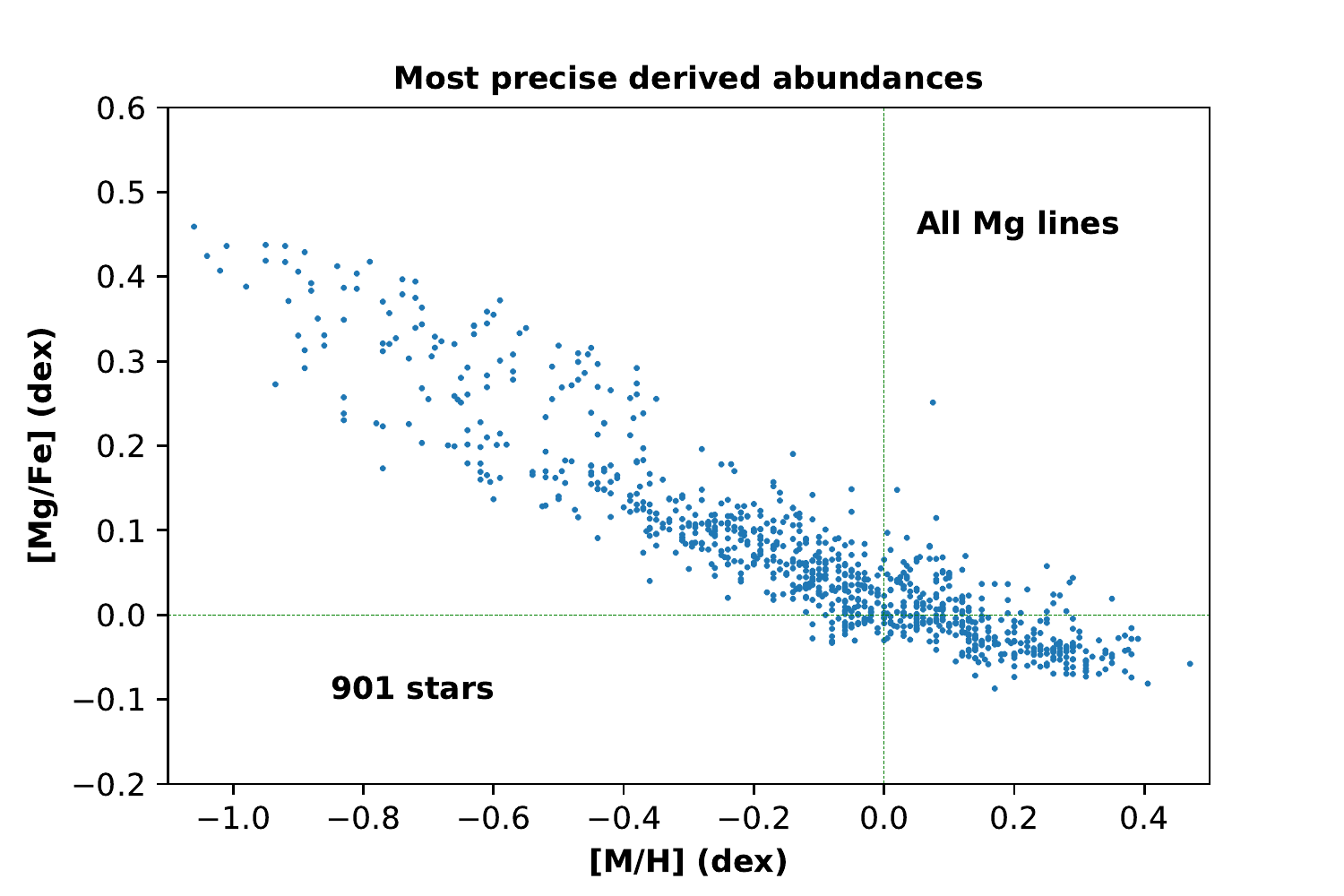}
\hspace{-0.7cm}
\includegraphics [height=70mm, width=0.45\textwidth] {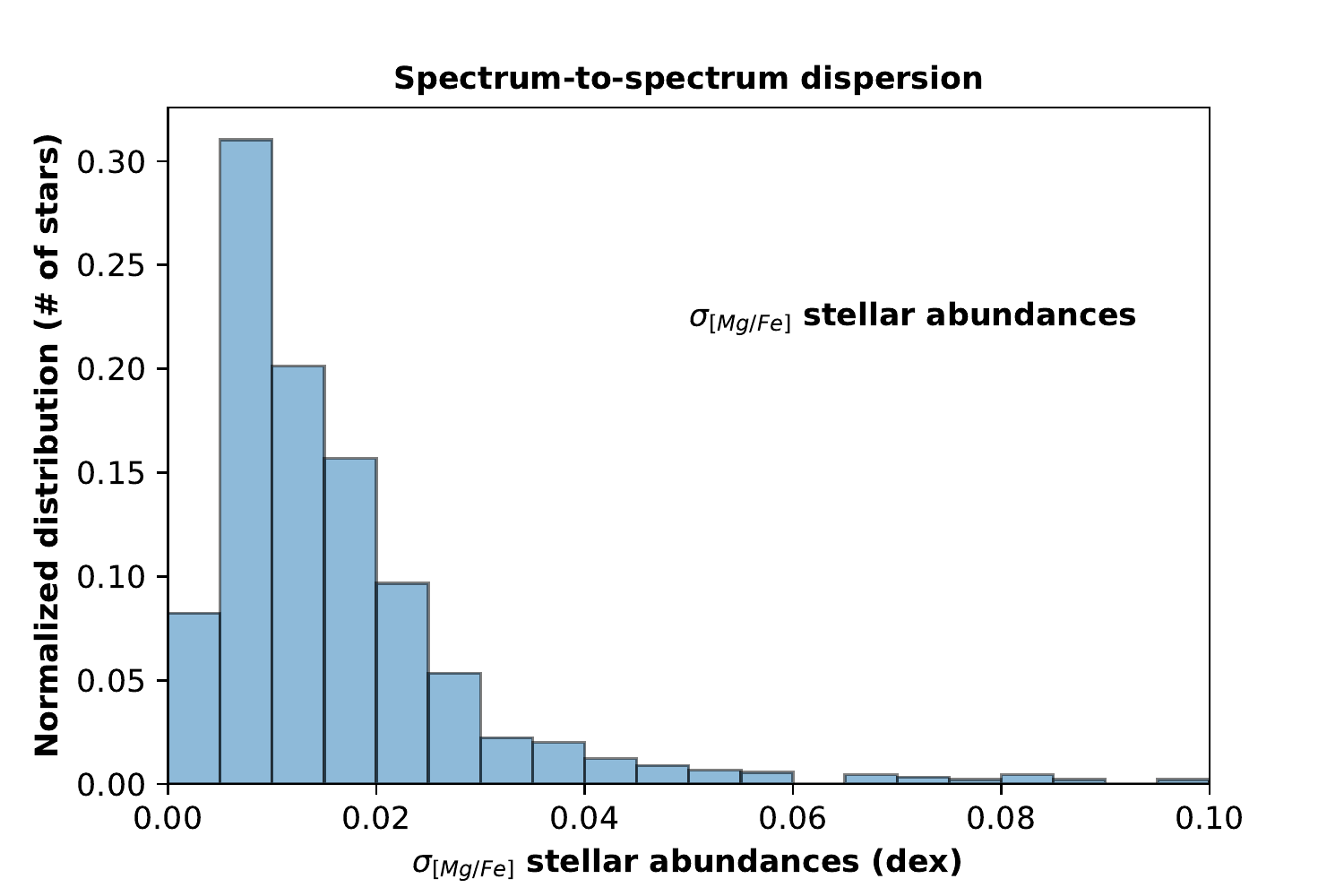}
\caption{\textbf{Left:} stellar abundance ratios [Mg/Fe] vs. [M/H] considering the abundance information from all the studied Mg I spectral lines. \textbf{Right:} estimated dispersion of the final stellar sample ($\geq$ 4 repeats) on the derived [Mg/Fe] abundances.} 
\label{Fig:dispersion_mejorCHI2}
\end{figure*}

\begin{figure*}
\centering
\includegraphics [height=65mm, width=0.95\textwidth] {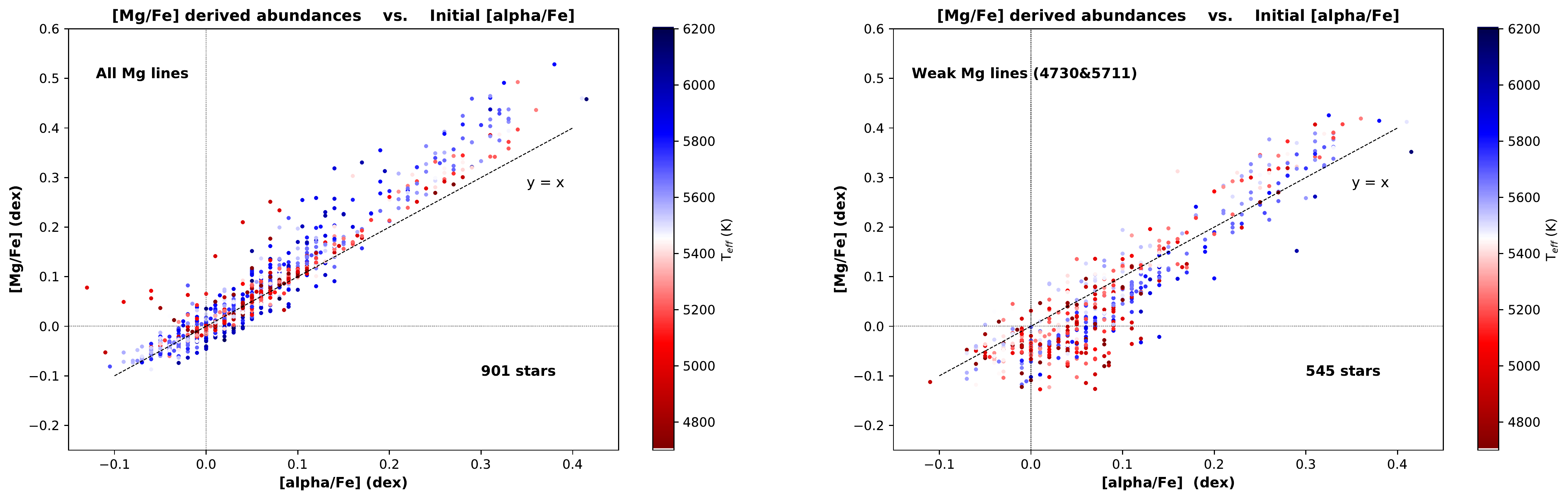}
\caption{Derived stellar abundance ratios [Mg/Fe] vs. initial stellar parameter [$\alpha$/Fe], colour-coded by stellar effective temperature. \textbf{Left:} considering the abundance information from all the studied Mg I spectral lines. \textbf{Right:} considering only the weak non-saturated lines 4730.04 $\&$ 5711.09 $\AA$ \citep[used in][as the input parameter to the strong saturated lines]{fuhrmann2017}.}
\label{Fig:final_MgvsAlpha}
\end{figure*}

The results for each Mg I line are presented separately in Fig. \ref{Fig:lines_mejorchi2}. The four saturated lines (MgIb triplet: 5167.3, 5172.7 \& 5183.6, and 5528.4 $\AA$) and the intermediate-strength line 5711.09$\AA$ seem to reproduce the thin-thick disc sequences more precisely, also showing a decreasing trend in [Mg/Fe] even at supersolar metallicities. This is in agreement with the analysis of NLTE effects on Mg abundances done by \citet{bergemann2017}, where they highlight the robust behaviour of the strong lines 5172, 5183, 5528, and 5711 $\AA$. The higher dispersion present on the abundance results for the weak non-saturated lines with respect to the strong saturated ones is due to different factors. On the one hand, for certain stellar types (towards hot metal-poor stars) and for lower Mg abundances (in terms of [Mg/H]), weak lines are closer to the spectral noise level. On the other hand, although saturated lines are less sensitive, pixel per pixel, to abundance variations (therefore presenting smaller flux variations per pixel), they span larger wavelength domains than non-saturated lines. As a consequence, the cumulative quantity of information on the abundance through all the considered pixels is very significant for strong lines, favouring a higher precision. This is confirmed by the study of the internal errors, through simulated noised theoretical spectra (see Appendix \ref{R_strong}). However, a high spectral resolution is required, even at high signal-to-noise values, to compensate the limited sensitivity of the lines to the abundance (c.f. Fig. \ref{Fig:SNR}). \par

Figure~\ref{Fig:dispersion_mejorCHI2} illustrates the final stellar abundance ratios [Mg/Fe] relative to [M/H] for our selected stellar sample (left panel), along with their estimated dispersion from the repeated observed spectra (right panel). The chemical distinction between the Galactic thin-thick disc populations is clearly observed, and the trend in [Mg/Fe] abundances at high metallicity ([M/H] > 0 dex) does not flatten. In addition, the gap in the [Mg/Fe]-enhanced disc stellar population, first observed by \citet{vardan2012} and later confirmed by \citet{sarunas2017}, still seems to be present in Fig.~\ref{Fig:lines_mejorchi2} \& Fig.~\ref{Fig:dispersion_mejorCHI2} around [M/H] $\approx$ -0.3 dex and [Mg/Fe] $\approx$ +0.2 dex. \par

Recent studies analysing dwarf stars in the solar neighbourhood \citep{sarunas2017, fuhrmann2017} present Mg lines that are in common with our work (c.f. Table 1 and Table 2 of \citealt{sarunas2017} and \citealt{fuhrmann1997}, respectively). \citet{sarunas2017} observed a different trend at high metallicities. However, as described before, \citet{sarunas2017} did not optimise the continuum normalisation for different stellar types, applying a constant local interval depending on the intensity of the line. On the other side, the agreement of our results with those of \citet{fuhrmann2017} is higher. Nevertheless, the slope of the low-alpha sequence in \citet{fuhrmann2017} seems less pronounced than in our work. \citet{fuhrmann2017} used the Mg I abundance estimate from weak lines as the input parameter to the Mg Ib lines (strong saturated). This first-guess abundance could have an important influence on the pseudo-continuum estimate, and therefore in the derived [Mg/Fe] abundance, for saturated lines. For cool or metal-rich dwarf stars, the impact of this first guess in the derived Mg Ib abundances could be stronger than for the other stars in the sample. As a consequence, possible parameter-dependencies in the results could remain. \par

Finally, we checked the consistency of our first-guess assumptions by comparing (on the left panel of Fig.~\ref{Fig:final_MgvsAlpha}) our initial abundance guess, coming from the [$\alpha$/Fe], and our final derived [Mg/Fe] abundances. The very good agreement between both quantities confirms the consistency of the procedure. In addition, the right panel of Fig.~\ref{Fig:final_MgvsAlpha} compares our initial guess with other suggested initial input in the literature, the abundance result from the weak lines (4730.04 and 5711.09 $\AA$) used by \citet{fuhrmann2017}. There is a very good agreement between our initial guess (the global [$\alpha$/Fe]) and the [Mg/Fe] from the weak lines, with only a few cool stars with [$\alpha$/Fe] around 0.1 dex with lower [Mg/Fe]$_{weak lines}$ values. Those few cases correspond to cool stars with supersolar metallicities. This suggests that the two weak lines could suffer from blends in the very crowded spectra, affecting the continuum placement and therefore the abundance estimate. As we can indeed see in Fig.~\ref{Fig:lines_mejorchi2}, the results of the two weak lines are more dispersed. On the other hand, the global [$\alpha$/Fe] parameter value was derived considering the complete HARPS wavelength domain \citep{DePascale2014}. Therefore, it is also less affected by continuum placement problems. For this reason, we believe that the global [$\alpha$/Fe] determined by the AMBRE pipeline, although very similar to the results of the weak lines, is a more precise initial guess for our application. \par

Our analysis allows a remarkable improvement with regard to previous efforts to chemically disentangle the Galactic thin-thick disc populations, and emphasises the importance of the normalisation procedure to properly interpret the chemical evolution of the disc. In conclusion, the feasibility of an optimal treatment on strong saturated lines to derive precise non-parameter-dependent abundances represents a major advancement. It will allow us to appropriately study the chemical signatures in the Galactic stellar populations and the resulting implications on chemodynamical relations, such as the abundance ratio [Mg/Fe] as a good age proxy, or the contribution of radial migration in the solar neighbourhood (Santos-Peral, in prep.).

%__________________________________________________________________

\section{Conclusions} \label{conclusions}
   
We carried out a detailed spectroscopic analysis of the Mg abundance estimation over a sample of 2210 FGK-type stars in the solar neighbourhood observed and parametrised at high spectral resolution (R=115000) within the context of the AMBRE Project. From this sample, we selected 1172 stars that have more than four observed spectra ($\geq$ 4 repeats). \par

We explored the possible sources of uncertainties in deriving chemical abundances, focusing on the continuum normalisation. From different stellar populations and nine Mg I spectral lines in the optical range, we observed different behaviours depending on the stellar type and the intensity of the line. \par

The normalisation procedure has an important impact on the derived abundances, with a strong dependence on the stellar parameters (T$_{eff}$, log g, [M/H]). Contrary to what is currently done in large spectroscopic surveys, the continuum placement procedure therefore has to be optimised for each stellar type and each spectral line. As expected, the intensity of the spectral lines has a drastic influence in the optimal width of the normalisation interval: 

\begin{itemize}
    \item \textbf{Non-saturated lines:} the optimal wavelength domain for the local continuum placement could be evaluated using a goodness-of-fit criterion, allowing a wavelength dependence with the spectral type. It is generally convenient to optimise the normalisation window close to the considered line (around two to four times their FWHM).  For strong (although not saturated) lines like 5711.09 $\AA$, a larger interval could be necessary when dealing with cool metal-rich stars ([M/H] > -0.2 dex ; T$_{eff}$ < 5400 K).
    \vspace{0.2cm}    
    \item \textbf{Saturated lines:} no pixels are available at the continuum level for most of the stellar types in the analysed region, and a pseudo-continuum normalisation has to be performed for the automatic fit. The level of the pseudo-continuum depends on the Mg abundance itself, which is at first assumed to be equal to the global $\alpha$-element abundance (or to another first-guess abundance). The induced bias in the Mg estimate is only partially corrected by the Mg line fitting due to a degeneracy between the fitting quality and the continuum placement (as the line is saturated, only the wings profile changes). In addition, the bias correction depends on the pseudo-continuum level itself. In this situation, one possibility consists in reducing the analysis of the saturated lines to the stellar types presenting continuum in the spectra (hot metal-poor stars). Alternatively, we have demonstrated that using a narrow normalisation window (around two to four times the FWHM of each line in a solar-type star) drastically reduces the parameter-dependence of the abundance estimate, increasing the line-to-line precision. This relies on the assumption that the Mg abundance behaviour is not very different from that of the global [$\alpha$/Fe] abundance.  
    
\end{itemize}

The final derived stellar abundance ratios [Mg/Fe], relative to [M/H], present a clear chemical distinction between the Galactic thin-thick disc populations and a decreasing trend in [Mg/Fe] abundances even at supersolar metallicities ([M/H] > 0 dex). With our analysis, we highlight the importance of carrying out an exploration on the optimisation of the continuum intervals in the preliminary and preparatory stages of any spectroscopic survey (in accordance with the stellar targets, spectral resolution, etc.). We suggest exploring the application of narrow normalisation windows around the studied lines in order to improve the abundance estimation precision. \par

The optimisation of the normalisation procedure in large spectroscopic stellar surveys would provide a significant improvement to the analysis of the chemical patterns of the different Galactic populations. The improvement in chemical abundance precision is strongly required in the present era of precise kinematical and dynamical data driven by the Gaia mission.

\begin{acknowledgements}
We would like to thank Nils Ryde and Mathias Schultheis for very useful suggestions and discussions. We thank the anonymous referee for his/her constructive comments, making a considerable contribution to the improvement of the paper. The authors thank M. Bergemann for providing her relation between the microturbulent velocity and the atmospheric parameters adopted for the spectra grid computation. This work is part of the PhD thesis project within the framework of "International Grants Programme" of the Instituto de Astrof\'isica de Canarias (IAC). P.S would like to thank the Centre National de Recherche Scientifique (CNRS) for the financial support. Part of this work was supported by the "Programme National de Physique Stellaire" (PNPS) of CNRS/INSU co-funded by CEA and CNES. A.R.B., P.dL. and E.F.A. acknowledge financial support from the ANR 14-CE33-014-01. E.F.A. also acknowledge partial support from the ANR 18-CE31-0017. This work has made use of data from the European Space Agency (ESA) mission {\it Gaia} (\url{https://www.cosmos.esa.int/gaia}), processed by the {\it Gaia} Data Processing and Analysis Consortium (DPAC, \url{https://www.cosmos.esa.int/web/gaia/dpac/consortium}). Funding for the DPAC has been provided by national institutions, in particular the institutions participating in the {\it Gaia} Multilateral Agreement. Most of the calculations have been performed with the high-performance computing facility SIGAMM, hosted by OCA.
\end{acknowledgements}

%-------------------------------------------------------------------

%%-----------------------------
%%   Bibliography
%%-----------------------------

%% The following lines are required when using BibTEX (strongly encouraged!):
\bibliographystyle{aa}  % A&A bibliography style file (aa.bst)
\bibliography{Santos-Peral} % your references in file: Yourfile.bib

\begin{appendix}

\section{Analysed local normalisation intervals} \label{INTERVALS}

\begin{figure*}
\centering
\includegraphics [height=180mm, width=0.45\textwidth] {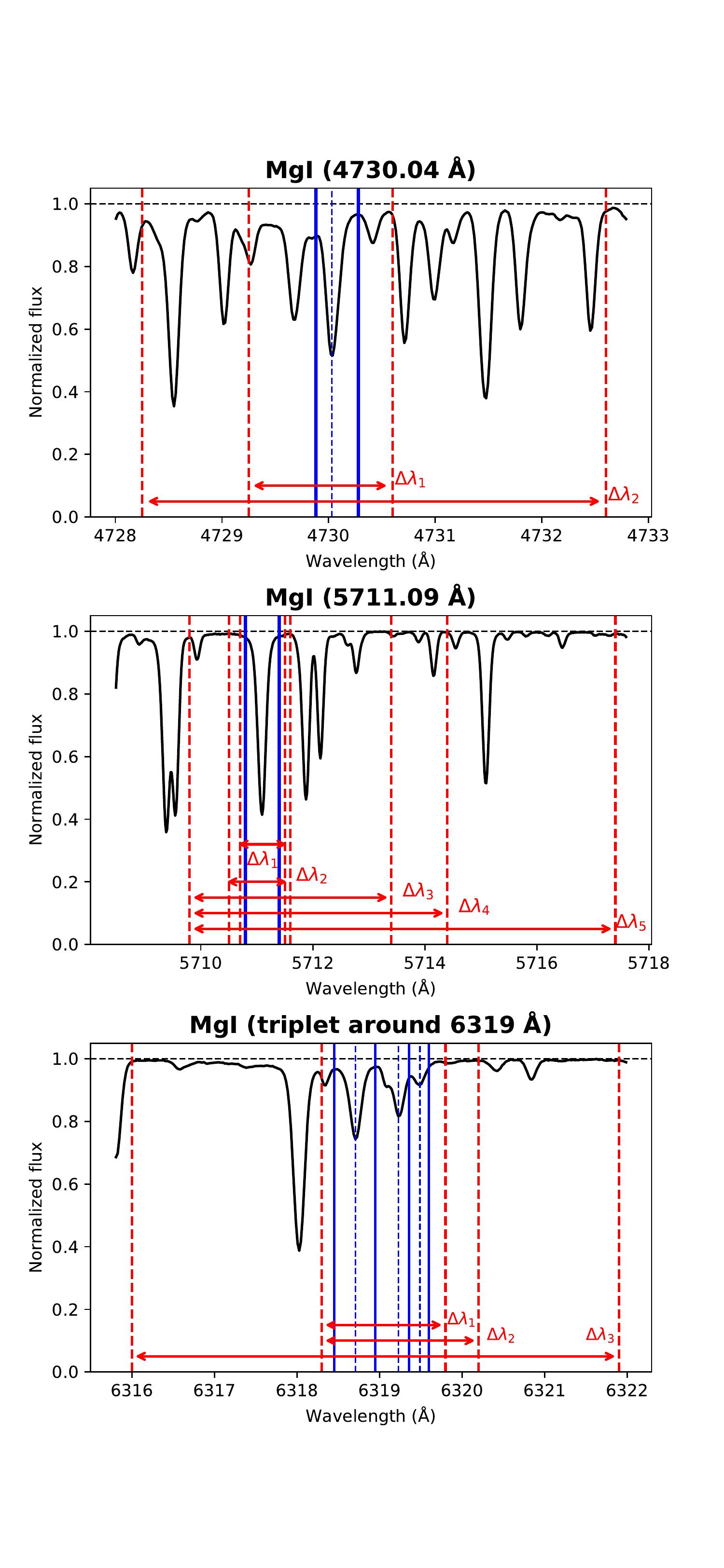}
\hspace{-0.5cm}
\includegraphics [height=200mm, width=0.48\textwidth] {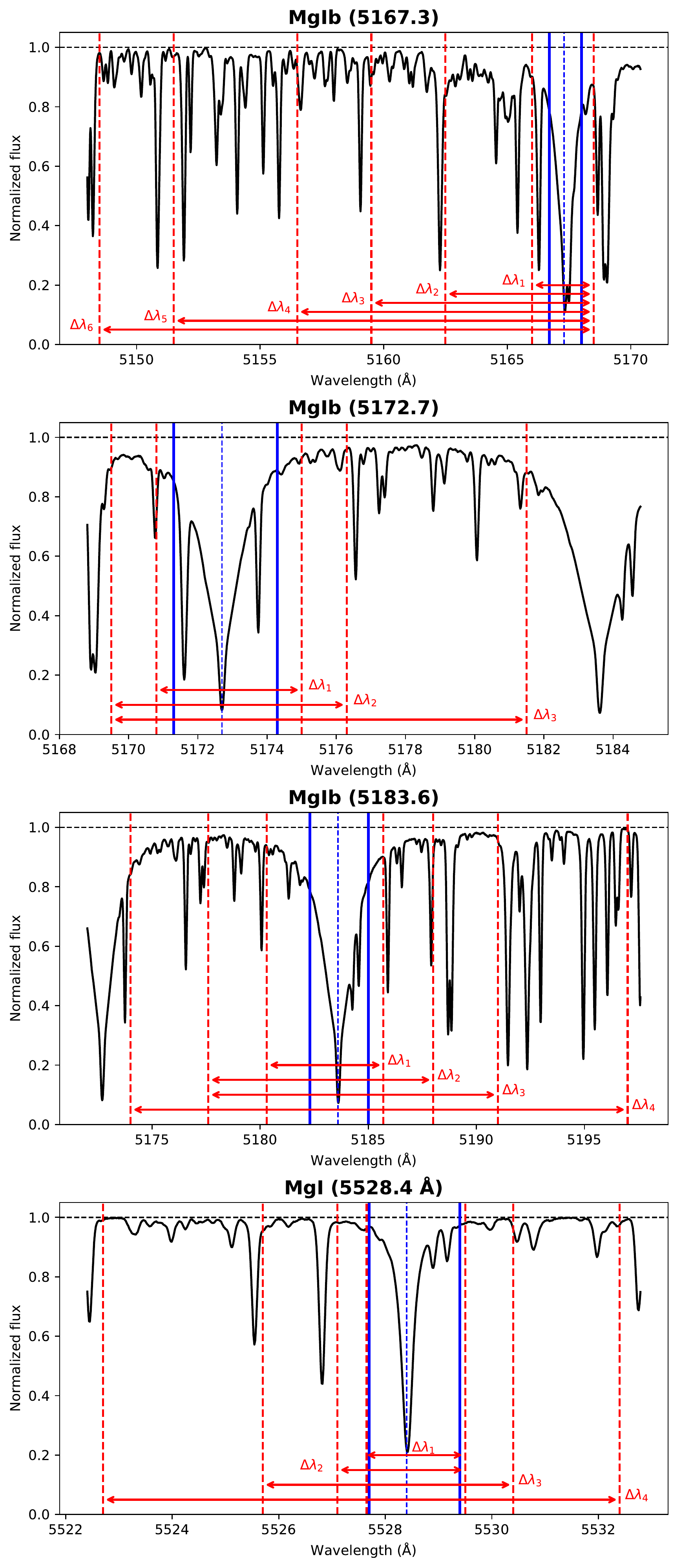}
\caption{Observed solar spectrum from HARPS around each Mg I spectral line. The abundance estimation window is delimited by blue vertical lines ($\Delta$$\lambda$$_{Abund}$ $\sim$ 0.5 $\AA$ for non-saturated lines and $\Delta$$\lambda$$_{Abund}$ $\sim$ 2.5 $\AA$ for strong saturated ones). The different local normalisation intervals applied in the analysis are shown with red dashed vertical lines. \textbf{Left:} non-saturated lines: 4730.04, 5711.09, 6318.7, 6319.24, and 6319.49 $\AA$. \textbf{Right:} strong saturated lines: 5167.3, 5172.7, 5183.6, and 5528.4 $\AA$ (from top to bottom).}
\label{Fig:intervals}
\end{figure*}

Different local continuum intervals were defined around each Mg I spectral line to explore their impact in the abundance analysis. By a spectral visualisation for different stellar types, the selection of the normalisation windows was adjusted to avoid the presence of strong absorption lines in the limits of the spanned region and chosen to have enough continuum points at the ends of the intervals.  Figure \ref{Fig:intervals} shows, for the observed solar spectrum, the different local normalisation intervals, along with the abundance estimation window, applied around each Mg line individually. In addition to the wavelength intervals plotted in this figure, we also tested a very large wavelength range around each analysed line  of $\Delta\lambda$ $\sim$ 70 $\AA$. The complete list of the defined normalisation windows around each line is shown in Table \ref{table:list}.

\begin{table*}
\centering
\begin{tabular}{ccccccc}
\\[\dimexpr-\normalbaselineskip+2pt]
\multicolumn{7}{c}{ \textbf{Normalisation windows ($\AA$)}} \\
\\[\dimexpr-\normalbaselineskip+2pt]
\hline
\hline
\\[\dimexpr-\normalbaselineskip+2pt]
\textbf{4730.04} & \textbf{5167.3} & \textbf{5172.7} & \textbf{5183.6} & \textbf{5528.4} & \textbf{5711.09} & \textbf{6319 triplet} \\
\\[\dimexpr-\normalbaselineskip+2pt]
\hline
\hline
\\[\dimexpr-\normalbaselineskip+2pt]
%\multicolumn{1}{c}{ \textbf{\emph{This work:}}} \\
 [4729.5,4730.6] & [5166.0,5168.5] & [5170.8,5175.0] & [5180.3,5185.7] & [5527.6,5529.5] & [5710.7,5711.5] & [6318.3,6319.8] \\
\\[\dimexpr-\normalbaselineskip+2pt]
[4728.2,4732.6] & [5162.5,5168.5] & [5169.5,5176.3] & [5177.6,5188.0] & [5527.1,5529.5] & [5710.5,5711.6] & [6318.3,6320.2] \\
\\[\dimexpr-\normalbaselineskip+2pt]
& [5159.5,5168.5] & [5169.5,5181.5] & [5177.6,5191.0] & [5525.7,5530.4] & [5709.8,5713.4] & [6316.0,6321.9] \\
\\[\dimexpr-\normalbaselineskip+2pt]
& [5156.5,5168.5] & & [5174.0,5197.0] & [5522.7,5532.4] & [5709.8,5714.4] \\
\\[\dimexpr-\normalbaselineskip+2pt]
 & [5151.5,5168.5] & & & & [5709.8,5717.4]\\
\\[\dimexpr-\normalbaselineskip+2pt]
 & [5148.5,5168.5] \\
\\[\dimexpr-\normalbaselineskip+2pt]
[4696 - 4764] & [5140 - 5210] & [5140 - 5210] & [5140 - 5210] & [5492 - 5564] & [5676 - 5746] & [6284 - 6354]\\
\hline
\hline
\end{tabular}
\vspace{0.10cm}
\caption{List of the selected local normalisation intervals around each Mg I spectral line.}
\label{table:list}
\end{table*}

%-----------------------------------------------------------------------------------------------------------------------------------------

\section{Sensitivity to spectral line-broadening} \label{rotation}

\begin{figure*}
\centering
\includegraphics [height=68mm, width=0.5\textwidth] {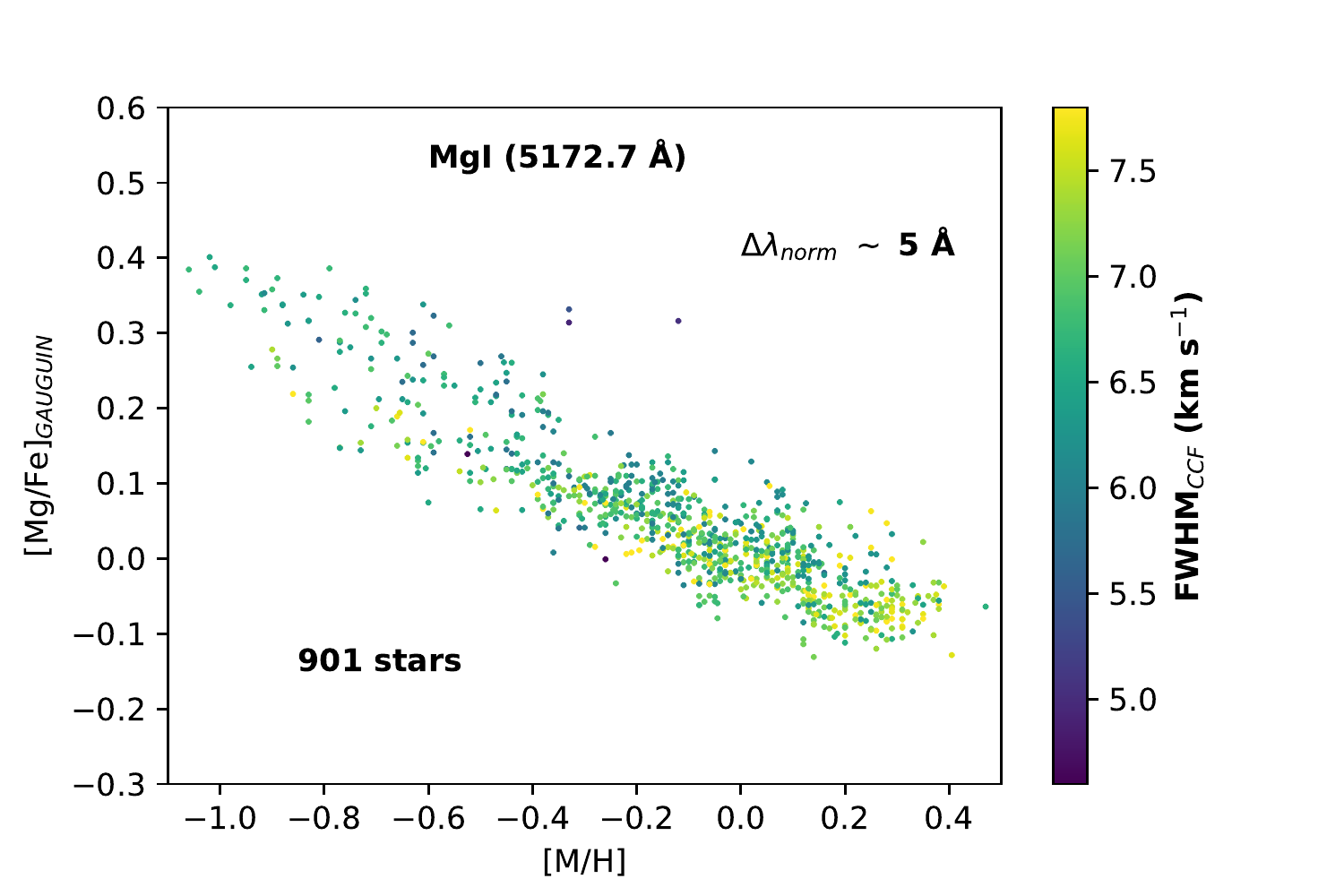}
\hspace{-0.5cm}
\includegraphics [height=68mm, width=0.5\textwidth] {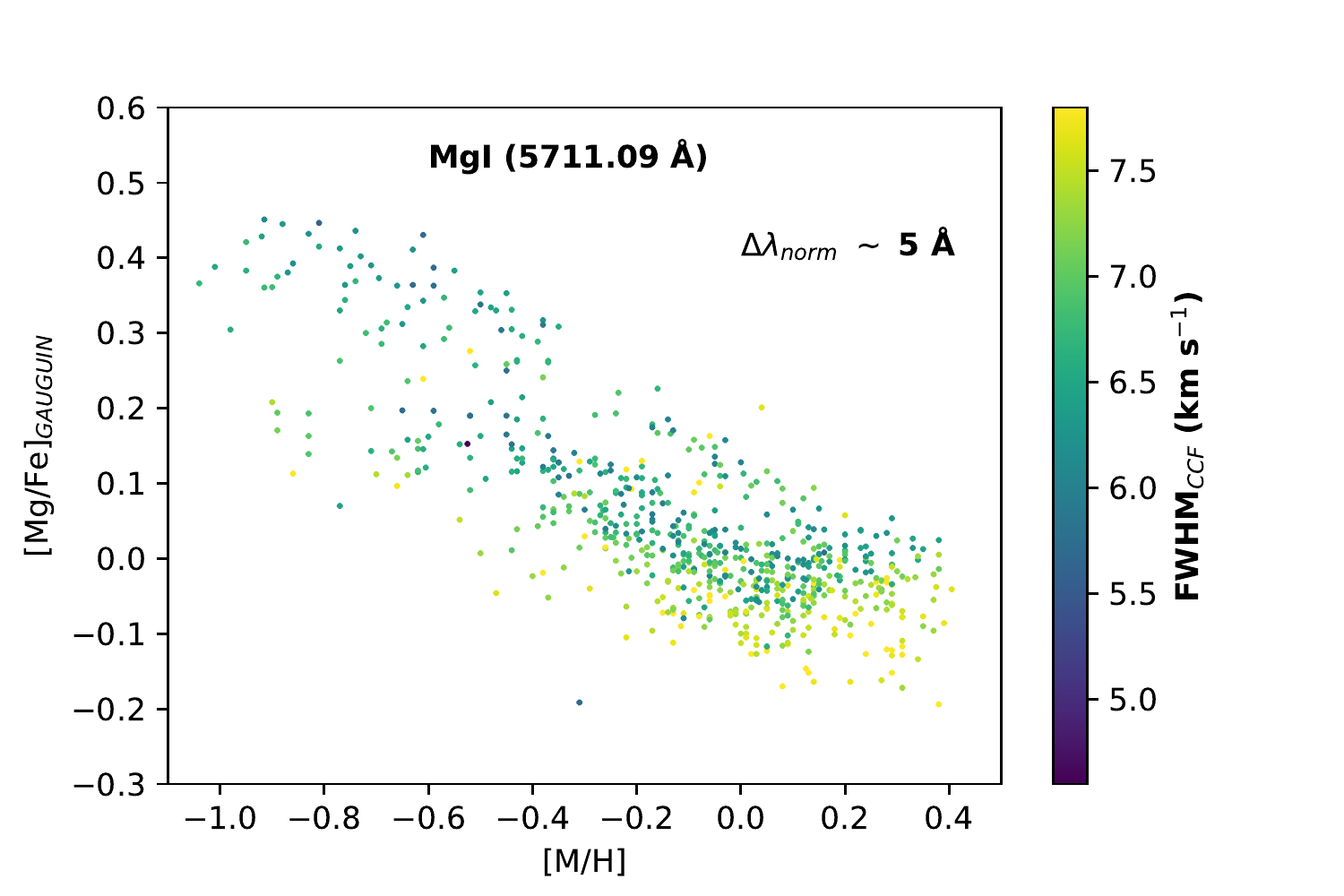}
\caption{Abundance ratio [Mg/Fe] as a function of [M/H], colour-coded by the FWHM$_{CCF}$ of the cross-correlation function. Both panels contain the same number of stars. Strong saturated lines (left, with less dispersed sequences) are less sensitive to spectral line-broadening than non-saturated lines (right). The [Mg/Fe] abundances were derived by performing the continuum placement around a local  wavelength interval of 5$\AA$.}
\label{Fig:rotation}
\end{figure*}

The AMBRE Project provides the FWHM of the cross-correlation function between the observed spectra and the corresponding templates used for the radial velocity estimation. This FWHM$_{CCF}$ can be used to study the sensitivity of the abundance precision to the line-broadening sources as stellar rotation and macroturbulence. \par

Figure \ref{Fig:rotation} shows the [Mg/Fe] vs. [M/H] abundances derived using two different spectral lines (see Sect \ref{lines}): a saturated one (5172.7$\AA$, left panel) and a non-saturated one (5711.09$\AA$, right panel). For each line, the continuum placement has been performed over a local window of the same width. We conclude that the dispersion on the [Mg/Fe] abundance measurement, with respect to [M/H], is dominated by the spectral line-broadening for non-saturated lines, although not for strong saturated lines. This is expected since the larger natural broadening of strong lines makes them less sensitive to the line-broadening. Similarly, high-resolution spectra are more sensitive to this effect than low-resolution data. This result highlights the relevance of a correct treatment when using weak non-saturated lines. Otherwise, choosing stronger lines or restricting the analysis to cool stars (for which the \textit{v$\sin$(i)} is lower) would minimise the effects. \par

We only kept spectra with FWHM$_{CCF}$ $\leq$ 8 km s$^{-1}$ for strong saturated lines, but only spectra with FWHM$_{CCF}$ $\leq$ 7 km s$^{-1}$ for non-saturated lines. As we do not have any \textit{v$\sin$(i)} determinations for these stars, we applied this cut based on the minimisation of the observed dispersion from each particular line.

%-----------------------------------------------------------------------------------------------------------------------------------------

\section{Effect of the spectral resolution on the abundance estimation for saturated lines} \label{R_strong}

\begin{figure*}
\centering
\includegraphics[height=55mm, width=\textwidth]{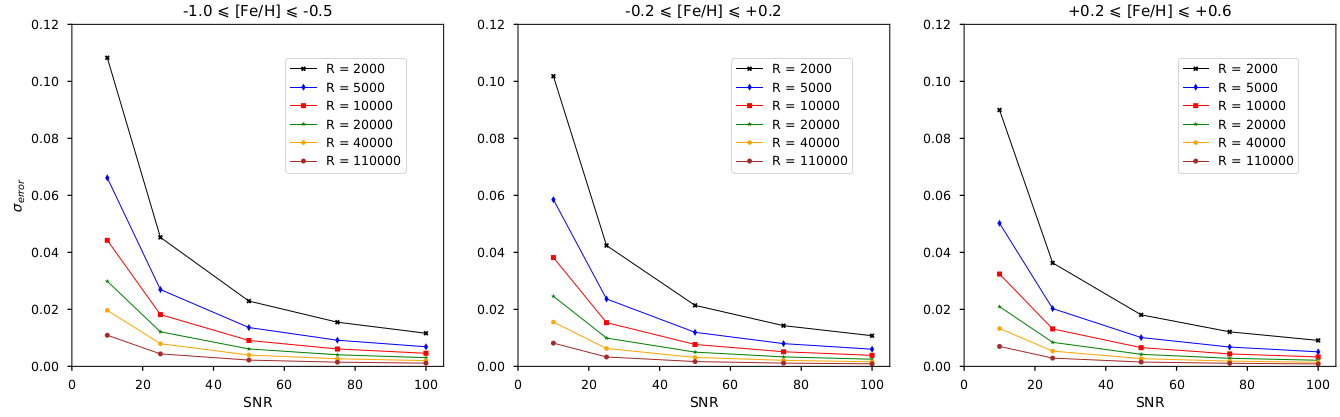}
\caption{Internal error, in the MgIb triplet lines, given as the standard deviation of the mean of the abundances difference ([Mg/Fe]$_{measured}$ - [Mg/Fe]$_{input}$) obtained from all measurements at a certain S/N and spectral resolution R, for a particular stellar type from the metal-poor (left) to the metal-rich (right) regime.}
\label{Fig:SNR}
\end{figure*}

We implemented a test with theoretical spectra to evaluate the internal error sources in our method of [Mg/Fe] estimation, concentrating on the strong lines of the MgIb triplet (5167.3, 5172.7, 5183.6 \AA). This error assessment leaves out the error sources restricted to real data, like the uncertainties in the line-broadening or the continuum normalisation, described in the body of the paper. In particular, the theoretical internal error analysis allows us to identify possible internal biases. \par

We built statistically significant sets of interpolated spectra from the high-resolution synthetic spectra grid, convolving to six different spectral resolutions, in the range from R $\sim$ 2000 to R $\sim$ 110000 (the HARPS resolution), and artificially adding random Gaussian noise (SNR = 10, 25, 50, 75, and 100). We defined four different stellar types; 
\begin{itemize}
    \item cool dwarf (4000 $\leq$ T$_{eff}$ $\leq$ 5000K, 4.4$\leq$ log(g) $\leq$5.0 cm s$^{-2}$)
    \item turn-off (5800 $\leq$ T$_{eff}$ $\leq$ 6200K, 3.5$\leq$ log(g) $\leq$ 4.1 cm s$^{-2}$)
    \item solar-type (5500 $\leq$ T$_{eff}$ $\leq$ 6000K, 4.2$\leq$ log(g) $\leq$ 4.7 cm s$^{-2}$)
    \item red clump (4000 $\leq$ T$_{eff}$ $\leq$ 4700K, 2.0$\leq$ log(g) $\leq$ 3.0 cm s$^{-2}$). 
\end{itemize}

For each case, at a given spectral resolution, we generated 250000 spectra (50000 per bin of S/N) for three different bins of metallicity: -1.0 $\leq$ [M/H] $\leq$ -0.5, -0.2 $\leq$ [M/H] $\leq$ +0.2, and +0.2 $\leq$ [M/H] $\leq$ +0.6 dex, performing the abundance estimation with GAUGUIN. \par

\begin{figure}
\centering
\includegraphics[height=50mm, width=0.5\textwidth]{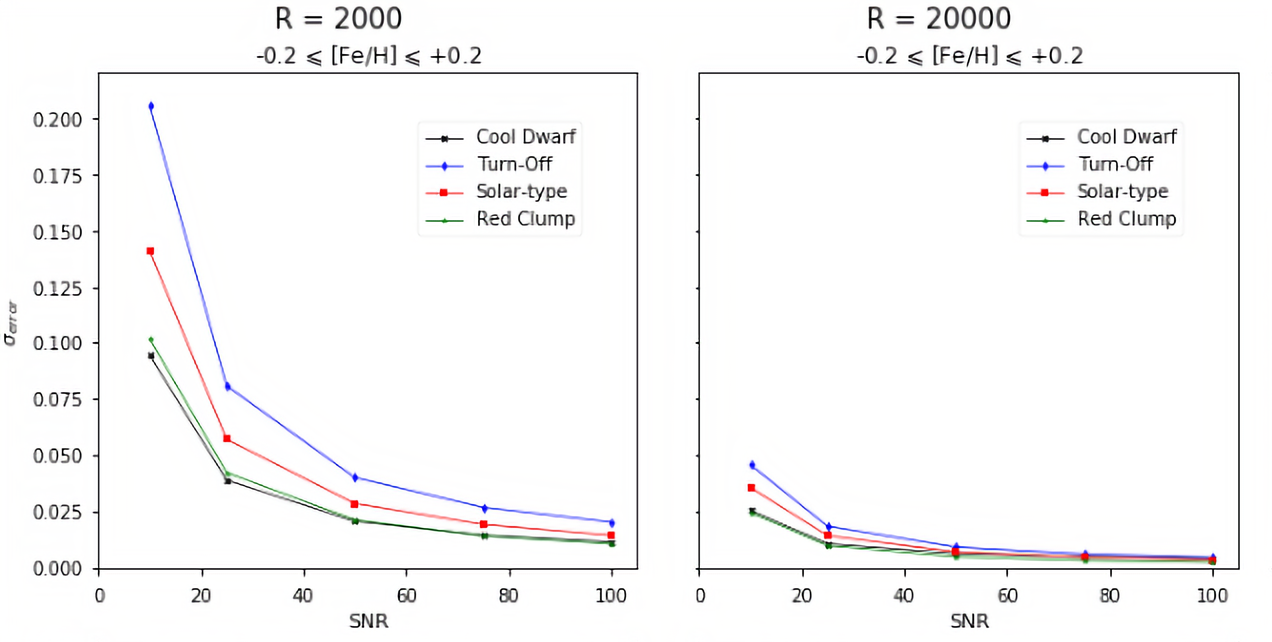}
\caption{Same as Fig. \ref{Fig:SNR} for different stellar types at low (left: R = 2000) and high (right: R = 20000) spectral resolution and solar metallicity.}
\label{Fig:R}
\end{figure}

First of all, no biases with stellar parameters are present in the results regardless of the resolution or signal-to-noise ratio of the data. Secondly, it appears that working at high resolution always leads to better results, independently of the stellar type and the metallicity range. In Fig. \ref{Fig:SNR}, the errors in the average [Mg/Fe] estimate (among the three MgIb lines) are given as a function of the SNR and for different resolutions. The three panels correspond to the three metallicity intervals. In Fig. \ref{Fig:R}, the abundance error is shown for the four considered stellar types (cool dwarf, turn-off, solar-type and red clump) at solar metallicity and changing the spectral resolution from 2000 (left) to 20000 (right). \par

These two figures illustrate that the abundance estimation precision depends principally on the spectral resolution and secondly on the target's effective temperature and metallicity. On the one hand, despite the fact that the considered lines are very large, the uncertainty increases critically when the spectral resolution is degradated. Although, as expected, this effect is more significant at low signal-to-noise values, increasing the SNR does not always compensate a resolution loss. In addition, even at high signal to noise, significant dependences of the uncertainty with the stellar type appear, if the spectral resolution is not high enough. In fact, the abundance uncertainties observed at low resolution are more dependent on the stellar type than those obtained at high resolution. This highlights the importance of working at high resolution for spectroscopic surveys targeting a variety of stellar types and metallicity ranges, in order to achieve more precise and homogeneous results, even when strong lines are used. Moreover, the pseudo-continuum is expected to become more significant (fewer pixels close to the continuum level) at lower spectral resolutions. \par

These results are in agreement with the analysis of \citet{nissen2018} regarding high-precision stellar abundances, underlining the relevance of carefully balancing the need for a large sample of stars against the spectral resolution and the S/N necessary to achieve a good precision in abundances.

\end{appendix}

\end{document}